\documentclass[letterpaper,11pt]{article}
\pdfoutput=1 

\usepackage{jheppub} 

\usepackage{graphicx}
\usepackage{epstopdf}
\usepackage{amsmath, amssymb}

\usepackage{tikz}

\usepackage{verbatim}
\usepackage{bm}
\usepackage{bbold}
\usepackage{caption}
\usepackage{subcaption}

\newcommand{\be}{\begin{eqnarray}}
\newcommand{\ee}{\end{eqnarray}}
\newcommand{\bea}{\begin{eqnarray}}
\newcommand{\eea}{\end{eqnarray}}

\newcommand{\bn}{\begin{enumerate}}
\newcommand{\en}{\end{enumerate}}

\def\IC{\mathbb{C}}

\def\IZ{\mathbb{Z}}

\def\CB{{\cal B}}
\def\CC{{\cal C}}

\def\CF{{\cal F}}

\def\CL{{\cal L}}

\def\CN{{\cal N}}

\def\CS{{\cal S}}
\def\CT{{\cal T}}

\def\half{\frac{1}{2}}

\def\Tr{{\rm Tr}}
\def\tr{{\rm Tr}}
\def\det{{\rm det}}

\title{Chiral theories of class $\CS$}

\author{Amihay Hanany}
\author{and Kazunobu Maruyoshi}

\affiliation{Imperial College London, Blackett Laboratory, Prince Concert Rd, London, SW7 2AZ, UK}

\emailAdd{a.hanany@imperial.ac.uk}
\emailAdd{k.maruyoshi@imperial.ac.uk}

\preprint{IMPERIAL-TP-15-AH-02}

\abstract
{We study a class of four-dimensional $\CN=1$ superconformal field theories 
obtained from the six-dimensional $(1,0)$ theory, on M5-branes on $\mathbb{C}^{2}/\mathbb{Z}_{k}$
orbifold singularity, compactified on a Riemann surface.
This produces various quiver gauge theories whose matter contents are chiral.
We classify the building blocks associated to pairs-of-pants,
and study the gauging of them as the gluing of punctures.
The Riemann surface picture makes the duality invariance of the resulting quiver theories manifest:
the theories associated to the same Riemann surface flow to the same nontrivial infrared fixed point.
We explicitly check this from the 't Hooft anomalies of the global symmetries and central charges.
}

\begin{document} 
\maketitle
\flushbottom


\section{Introduction}
  
  String/M-theory is a powerful tool to study various properties of supersymmetric gauge theories 
  on worldvolumes of branes.
  One of the remarkable discoveries is four-dimensional $\CN=2$ superconformal field theories (SCFTs)
  obtained from the M5-branes (or six-dimensional $\CN=(2,0)$ theory) compactification 
  on a Riemann surface \cite{Gaiotto:2009we}.
  The construction produces a class of theories, so called class $\CS$, 
  involving various building blocks associated to pairs-of-pants, 
  which are free theories of hypermultiplets or interacting SCFTs.
  This framework allows us to uniformly understand non-perturbative properties {\it e.g.}
  the S-dualities, of theories in this class in terms of the Riemann surface.
  
  Generalizations to $\CN=1$ superconformal theories and their low energy physics have been studied 
  in \cite{Maruyoshi:2009uk,Benini:2009mz,Bah:2011je,Bah:2012dg,Gadde:2013fma,Xie:2013gma,Bah:2013aha,
  Agarwal:2013uga,Agarwal:2014rua,McGrane:2014pma}. 
  As in the $\CN=2$ case, (four-dimensional) UV descriptions of this class of theories consist 
  of {\it non-chiral} building blocks,
  like a pair of fundamental and anti-fundamental chiral multiplets.
  Therefore this is a small (but of course interesting) subset of possible $\CN=1$ superconformal theories.
  
  An important step is, thus, to incorporate {\it chiral}-ness to this kind of construction.
  As can be seen from the earlier works in \cite{Lykken:1997gy,Uranga:1998vf,Oh:1999sk}, 
  one way to achieve this is to consider the orbifold acting on the transverse directions to M5-branes.
  Namely, $N$ M5-branes on the $\mathbb{C}^{2}/\mathbb{Z}_{k}$ orbifold singularity
  whose worldvolume theory is a six-dimensional $\CN=(1,0)$ theory 
  \cite{Intriligator:1997kq,Blum:1997mm,Brunner:1997gf,Hanany:1997gh,DelZotto:2014hpa}, 
  compactified on a Riemann surface.
  A systematic study of this construction was recently done in \cite{Gaiotto:2015usa}.
  The four-dimensional quiver gauge theories obtained by the compactification are roughly 
  the orbifolded version of $\CN=2$ class $\CS$ theories,
  where an $\CN=2$ vector and hypermultiplets decompose into a number of $\CN=1$ $SU(N)$ vector multiplets connected 
  by $\CN=1$ bifundamental chiral multiplets, and a number of $\CN=1$ chiral multiplets respectively.
  Each $SU(N)$ gauge group effectively has $3N$ sets of fundamental and anti-fundamental chiral multiplets.
  (See also \cite{Ohmori:2015pua,DelZotto:2015rca} for recent works on the torus compactification 
  of six-dimensional $(1,0)$ theories.)
  
  In this paper we study a wider class of $\CN=1$ chiral theories 
  where each $SU(N)$ gauge group can have lower number of the flavors,
  and flow to nontrivial infrared fixed point. 
  From the point of view of the six-dimensional $\CN=(1,0)$ theory, this is obtained by 
  the introduction of a curvature of the particular global $U(1)_{t}$ symmetry on the Riemann surface,
  and this corresponds, roughly, to the orbifolded version of $\CN=1$ class $\CS$ theories described above.
  
  We find that the building blocks consisting of bifundamental chiral multiplets 
  are associated to a pair-of-pants with two maximal and one minimal punctures
  with additional information coming from the curvature.
  A maximal puncture has an $SU(N)^{k}$ flavor symmetry, and the gluing of these punctures
  corresponds to the gauging of this symmetry.
  The construction automatically ensures anomaly-free-ness of the gauge groups. 
  This leads, therefore, to consistent quiver theories associated to a cylinder or a torus,
  which is indeed the surface which we compactify the six-dimensional theory on.
  
  The geometric origin of this class of theories can be seen from
  the 't Hooft anomalies of the global symmetries, as they are written in terms of the compactified Riemann surface,
  and the central charges.
  A nice property of the central charges is that 
  the ``canonical'' $U(1)_{R}$ symmetry can only mix with the $U(1)_{t}$ mentioned above.
  Therefore this simplifis the a-maximization problem \cite{Intriligator:2003jj} 
  and gives a generic pattern of the mixing. 
  We also consider the cases where a mixing with another $U(1)$ symmetry takes place.
  
  We should mention that this kind of $\CN=1$ chiral theories are not new.
  Indeed, a large family of $\CN=1$ theories has been studied by brane tiling \cite{Franco:2005rj}
  as quivers on $T^{2}$.
  Generalization to quivers on a generic Riemann surface were studied 
  in \cite{Hanany:2012vc,Franco:2012mm,Xie:2012mr,Franco:2012wv,Cremonesi:2013aba}.
  This Riemann surface is closely related to (but not same as) 
  the one on which we compactify the six-dimensional theory,
  and the well-understood quantities in the brane tiling are translated to geometric objects studied here.
  For example some zig-zag paths on the former Riemann surface is related 
  to minimal punctures on the latter one.
  The brane tiling, however, includes quiver theories 
  which are not in the class constructed from the building blocks described in this paper.
  We will discuss possible ways toward these theories from the six-dimensional theory viewpoint.

  The organization of this paper is as follows:
  In Section \ref{sec:SQCD}, we start with the theory obtained by the orbifold projection 
  of the $\CN=1$ supersymmetric $SU(N)$ gauge theory 
  with $2N$ sets of fundamental and anti-fundamental chiral multiplets,
  which highlights the classification of the building blocks.
  We study various Seiberg-dual descriptions on this gauge theory.
  In Section \ref{sec:class}, the six-dimensional or string theory origin of the class of theories
  will be analyzed.
  We will see the generic construction of linear and cyclic quiver gauge theories using the building blocks. 
  In Section \ref{sec:higgs}, we consider the Higgsing by giving vevs to baryon operators.
  This corresponds to closing a minimal puncture with the introduction of another type of $U(1)$ curvature,
  and thus produces a new type of building blocks.
  In Section \ref{sec:anomaly}, we study the 't Hooft anomalies and central charges of the class of chiral theories.

  While completing this paper we received \cite{Franco:2015jna} 
  where the same quiver gauge theories were considered from the different perspective.

\section{Orbifolded SQCD and dualities}
\label{sec:SQCD}
  In this section we study the theory obtained by an orbifold projection on the four-dimensional $\CN=1$ 
  supersymmetric $SU(N)$ gauge theory with $2N$ sets of fundamental and anti-fundamental chiral multiplets, 
  with a quartic superpotential, including its dual theories.
  We will refer to the latter as $\CN=1$ SQCD with $N_{f} = 2N$.
  
\begin{figure}[t]
    \begin{center}
    \includegraphics[width=13cm, bb=0 0 857 189]{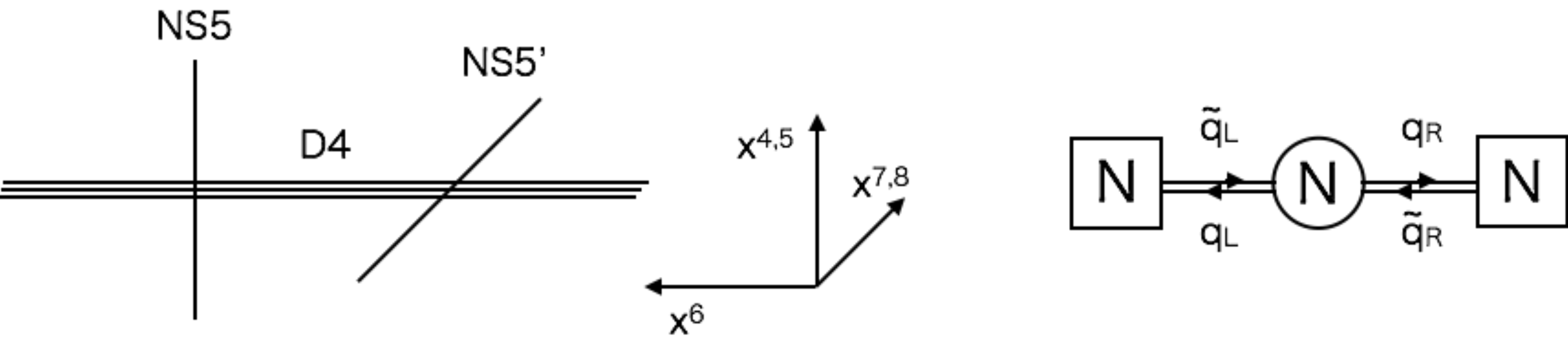}
    \caption{Left: Type IIA brane configuration of $\CN=1$ SQCD with $N_{f}=2N$ flavors.
             Right: its quiver diagram.}
    \label{fig:braneSQCD}
    \end{center}
    \end{figure}

  To specify the orbifold action, we first consider the Type IIA brane configuration 
  of the $\CN=1$ SQCD \cite{Elitzur:1997hc}.
  This is given by D4-, NS5- and NS5$'$-brane system which are occupying 
  the directions $x_{i}$ with $i=0,1,2,3,6$, $i=0,1,2,3,4,5$ and $i=0,1,2,3,7,8$, respectively,
  as depicted in Figure \ref{fig:braneSQCD}.
  The $N$ D4-branes suspended between NS5- and NS5$'$-branes give rise to $\CN=1$ $SU(N)$ vector multiplet
  and the $N$ D4-branes stretched to $+ \infty$ ($-\infty$) give rise to $N$ sets of fundamental and antifundmental 
  chiral multiplets, $q_{L}$ and $\tilde{q}_{L}$ ($q_{R}$ and $\tilde{q}_{R}$).
  We add to this theory a particular marginal superpotential coupling 
  $W= \tr (q_{L} \tilde{q}_{L})_{adj} (q_{R} \tilde{q}_{R})_{adj}$,
  where $(\ldots)_{adj}$ means the combination of $...$ 
  transforming in the adjoint representation of the gauge group,
  as introduced in \cite{Gadde:2013fma}.
  This preserves, of course, the $U(1)_{R}$ symmetry and another $U(1)_{\CF}$ symmetry whose geometric origin 
  are two combinations of the isometries of $v = x^{4} + i x^{5}$ and $w = x^{7}+ix^{8}$ planes.
  In particular $U(1)_{\CF} = \frac{U(1)_{v} - U(1)_{w}}{2}$,
  where the quarks $q_{L}$ ($q_{R}$) have charge $+\frac{1}{2}$ ($- \frac{1}{2}$) under $U(1)_{\CF}$.
  
  We now consider the orbifold $\IC^{2}/\IZ_{k}$ which acts on the coordinate as
    \bea
    v
     =     e^{2\pi i/k} v, ~~~~
    w
     =     e^{-2 \pi i/k} w.
    \eea
  (See \cite{Lykken:1997gy,Uranga:1998vf,Oh:1999sk} for the same orbifold of the various 
  $\CN=2$ and $\CN=1$ theories.)
  This orbifold group is a discrete subgroup of $U(1)_{\CF}$.
  To take an orbifold we change the gauge group and flavor groups to $SU(kN)$.
  Then we find a discrete subgroup of the $SU(kN)$ gauge group
  which acts on the fundamental representation as 
  ${\rm diag} (\mathbb{1}_{N}, \alpha \mathbb{1}_{N}, \alpha^{2} \mathbb{1}_{N},\ldots)$.
  We identify the diagonal part of this action and the $\mathbb{Z}_{k}$ subgroup of $U(1)_{\CF}$ with the orbifold
  and take the invariant part of them.
  It is easy to see that the vector multiplet simply decomposes into $\prod_{i=0}^{k-1} SU(N)_{i}$ groups.
  In a similar way for the (anti-)fundamental chiral multiplets, 
  we consider the subgroups which act on the fundamental representations as
  ${\rm diag} (\alpha^{-1/2} \mathbb{1}_{N}, \alpha^{1/2} \mathbb{1}_{N}, \alpha^{3/2} \mathbb{1}_{N},\ldots)$ 
  for the both flavor $SU(kN)$.
  For the left sets of chirals with $U(1)_{\CF}$ charge $\frac{1}{2}$, this projects out to
  $Q_{i}$ which transform in the bifundamental representations $(\bar{\bf{N}}_{i}, \bf{N}_{i-1}^{g})$
  and $\tilde{Q}_{i}$ transforming in $(\bf{N}_{i}, \bar{\bf{N}}_{i}^{g})$
  where $i=0,1,\ldots,k-1$ and $\bf{N}^{g}_{i}$ are the fundamental representations of the gauge symmetries. 
  For the right sets of chiral multiplets with $U(1)_{\CF}$ charge $-\frac{1}{2}$, 
  the projection is similar with the exchange of the fundamental and anti-fundamental representations.
  Thus we get $q_{i}$ and $\tilde{q}_{i}$ transforming in $(\bf{N}^{g}_{i}, \bar{\tilde{\bf{N}}}_{i})$ 
  and $(\bar{\bf{N}}_{i}^{g}, \tilde{\bf{N}}_{i+1})$.
  
\begin{figure}[t]
    \begin{center}
    \includegraphics[width=4.5cm, bb=0 0 294 434]{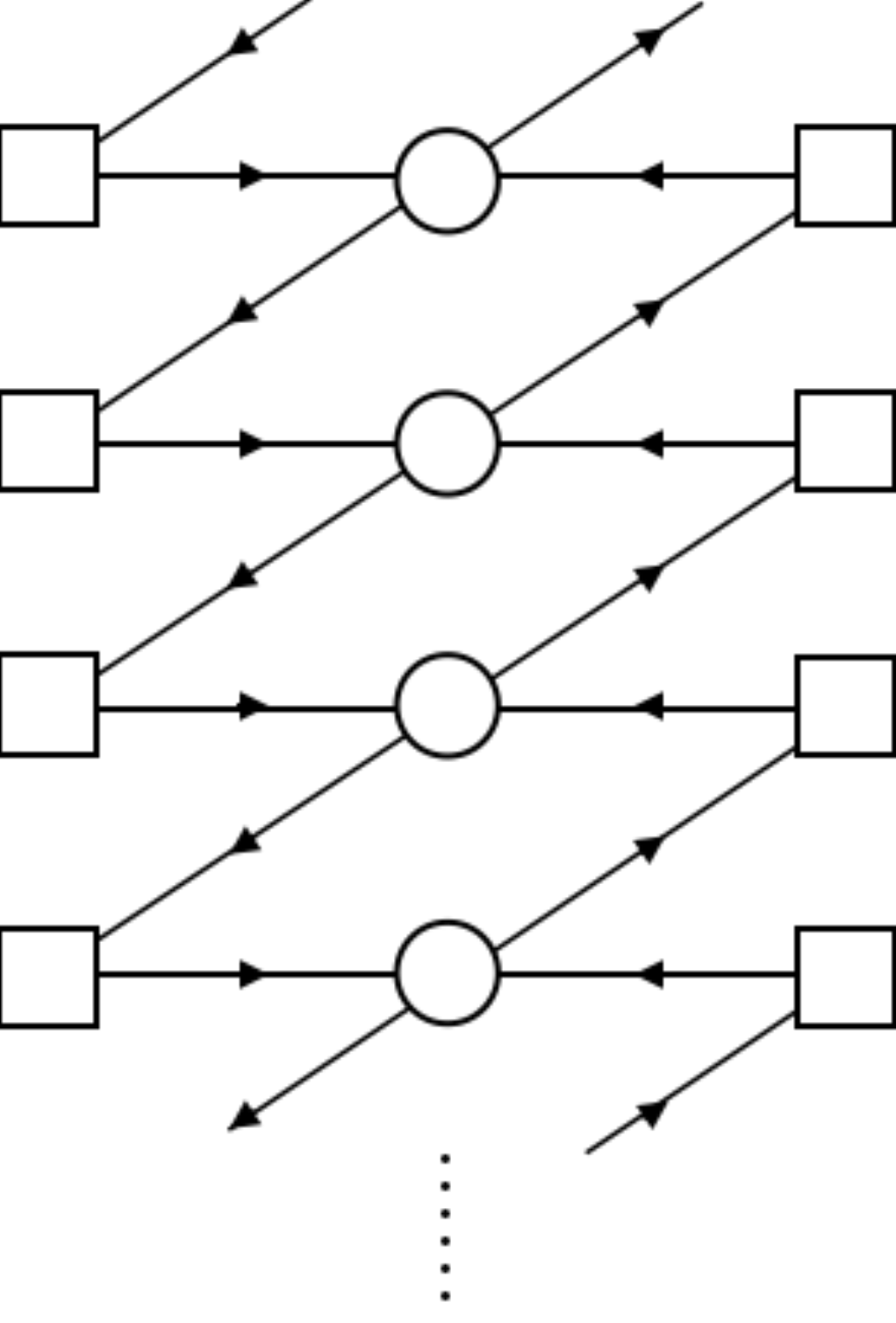}
    \caption{The orbifold projection of the $\CN=1$ SQCD with $N_{f}=2N$ flavors and the quartic coupling.
             A circle and a box represent an $\CN=1$ $SU(N)$ gauge and an $SU(N)$ flavor symmetries respectively.
             A line with an arrow represents an $\CN=1$ chiral multiplet in the bifundamental representation 
             of two groups.}
    \label{fig:SQCD1}
    \end{center}
    \end{figure}

  The quartic coupling is projected to the following terms
    \bea
    W
     =     \sum_{i=0}^{k-1} \tr Q_{i} \tilde{Q}_{i} q_{i} \tilde{q}_{i-1},
    \eea
  where $\tilde{q}_{-1}\equiv \tilde{q}_{k-1}$.
  The resulting theory is depicted in Figure \ref{fig:SQCD1},
  where the vertical direction is periodic and there are $k$ $SU(N)$ groups.
  For each oriented rhombus there is a quartic coupling.
  
  This theory is chiral, however free from the gauge anomaly: each $SU(N)$ gauge group has 
  $2N$ fundamental and $2N$ anti-fundamental chiral multiplets.
  In other words, this has the same matter content as that of the SQCD with $N_{f} = 2N$.
  This value is in the middle of the conformal window, thus we expect that the theory
  flows to the strongly coupled IR fixed point.
  
  We now consider the properties of this quiver theory and its infrared SCFT.
  
\subsection{Building blocks}
\begin{figure}[t]
	\begin{center}
	\includegraphics[width=13cm,bb=0 0 894 433]{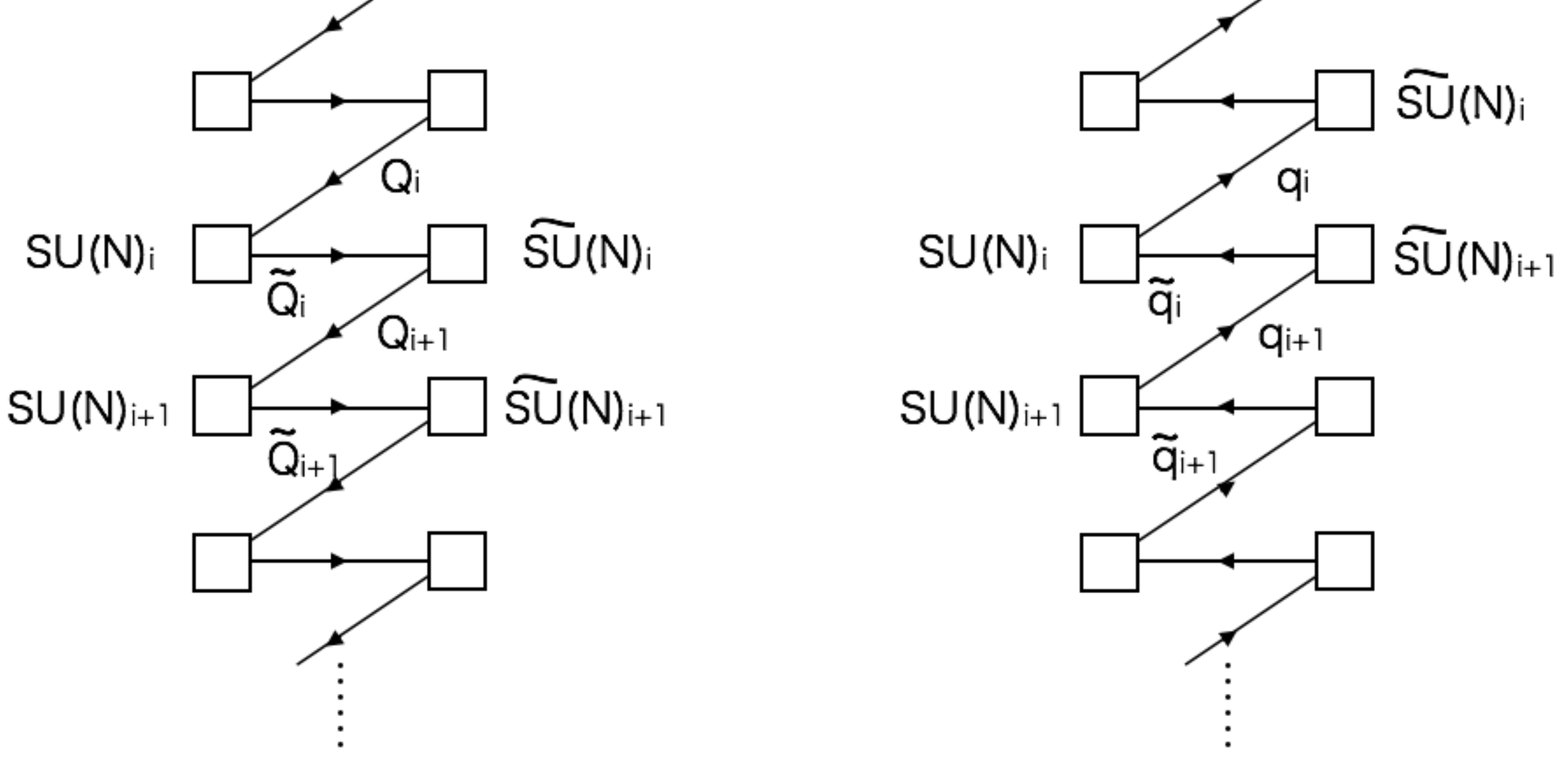}
	\caption{Graphical representation of a basic building block. 
	         We will refer to the left and the right building blocks 
	         as $\CB^{++}_{+}$ and $\CB^{--}_{-}$ respectively.}
	\label{fig:buildingblock1}
	\end{center}
    \end{figure}

  By decoupling all the gauge groups, we get two different building 
  blocks\footnote{From the IR theory point of view, we cannot decouple the gauge groups.
                  This can be seen from that the exactly marginal coupling does not continue to 
                  a weakly coupling regiem for gauge coupling.
                  Nevertheless the following argument is useful to read off the matter content 
                  of the four-dimensional UV theory.}, 
  as in Figure \ref{fig:buildingblock1}.
  Let us focus on the left one which consists of $2k$ chiral multiplets $Q_{i}$ and $\tilde{Q}_{i}$.
  In addition to the $SU(N)^{2k}$ symmetry, there are $2k$ $U(1)$ global symmetries, 
  which come from the symmetry rotating chiral multiplets, and $U(1)_{R}$ symmetry.
  We parametrize non-R $U(1)$'s 
  as $U(1)_{\alpha} \times U(1)_{t} \times \prod_{i=0}^{k-1} U(1)_{\beta_{i}} 
  \times \prod_{i=0}^{k-1} U(1)_{\gamma_{i}}$
  with the constraints that the sum of all $U(1)_{\beta_{i}}$ charges gives $U(1)_{t}$ up to a factor,
  and the same for $U(1)_{\gamma_{i}}$.
  We denote the assignment of the charges in the following notation:
    \bea
    Q_{i}: R^{\half} t^{\half} \beta_{i} \alpha, ~~~
    \tilde{Q}_{i}: R^{\half} t^{\half} \gamma_{i}^{-1} \alpha^{-1} \nonumber,
    \eea
  which means $Q_{i}$ has charge $\half$ both for $U(1)_{R}$ and $U(1)_{t}$, 
  $1$ for $U(1)_{\beta_{i}}$ and $1$ for $U(1)_{\alpha}$.
  
  We can represent this theory by specifying the 't Hooft anomalies among global symmetries
    \bea
    \tr R
    &=&  - kN^{2}, ~~~
    \tr R^{3}
     =   - \frac{kN^{2}}{4}, ~~~
    \tr T
     =     k N^{2}, ~~~
    \tr T^{3}
     =     \frac{kN^{2}}{4},  
           \nonumber \\
    \tr T R^{2}
    &=&    \frac{kN^{2}}{4}, ~~~
    \tr T^{2} R
     =   - \frac{k N^{2}}{4}, ~~~
    \tr \beta_{i}
     =     \tr \beta_{i}^{3}
     =   - \tr \gamma_{i}
     =   - \tr \gamma_{i}^{3}
     =     N^{2}, 
           \nonumber \\
    \tr R SU(N)_{i}^{2}
    &=&  - 2 \tr T SU(N)_{i}^{2}
     =     \tr R \widetilde{SU}(N)_{i}^{2}
     =   - 2 \tr T \widetilde{SU}(N)_{i}^{2}
     =   - \frac{N^{2}}{2}, 
           \nonumber \\
    \tr \beta_{i} SU(N)_{i}^{2}
    &=&  - \tr \gamma_{i} SU(N)_{i}^{2}
     =     \frac{N}{2}, ~~~
    \tr \beta_{i+1} \widetilde{SU}(N)_{i}^{2}
     =   - \tr \gamma_{i} \widetilde{SU}(N)_{i}^{2}
     =     \frac{N}{2},
           \nonumber \\
    \tr \beta_{i} \alpha^{2}
    &=&  - \tr \gamma_{i} \alpha^{2}
     =     N^{2},~~~
    \tr T \alpha^{2}
     =   - \tr R \alpha^{2}
     =     k N^{2},
           \label{betaanom1}
    \eea
  where $R$, $T$, $\alpha$, $\beta_{i}$ and $\gamma_{i}$ are the generators of the corresponding $U(1)$ symmetries 
  respectively\footnote{Here we are using a convention that the Casimir of the fundamental representation gives $1/2$.}.
  
  As will be clear in the next section, 
  the $U(1)_{R} \times U(1)_{t} \times U(1)_{\beta_{i}} \times U(1)_{\gamma_{i}}$ 
  is the ``intrinsic'' symmetry which originates from the six-dimensional theory.
  Therefore this symmetry exists in the class of theories obtained from the building block we are studying.
  The other symmetries come from the compactification of six-dimensional theory
  on the Riemann surface with punctures.
  The $SU(N)^{2k}$ global symmetry comes from the two maximal punctures.
  The puncture can be specified by the anomalies \cite{Gaiotto:2015usa}
    \bea
    \tr \beta_{i + n - o} SU(N)_{i}^{2}
     =     \frac{N}{2}, ~~~
    \tr \gamma_{i} SU(N)_{i}^{2}
     =   - \frac{N}{2},
    \eea
  where $n$ ($n=0,1,\ldots,k-1$) and $o$ (by convention, $o=+1$ for a left maximal puncture, and $o=-1$ for a right maximal puncture) are two parameters which are used to shift the label $i$. They are called a ``color'' and an ``orientation'' of the puncture, respectively.
  By using \eqref{betaanom1} we read off the color and the orientation of the punctures of the building blocks
  as $o=+1$ and $n=1$ for the left puncture and $o=-1$ and $n=0$ for the right puncture, respectively.
  The $U(1)_{\alpha}$ symmetry is a baryonic symmetry which is associated to the minimal puncture.

  We now consider another building block, depicted in Figure \ref{fig:buildingblock1}, which is specified by the anomaly coefficients
    \bea
    \tr \beta_{i + n - o} SU(N)_{i}^{2}
     =   - \frac{N}{2}, ~~~
    \tr \gamma_{i} SU(N)_{i}^{2}
     =     \frac{N}{2},
           \label{betaanom2}
    \eea
  which sets the labels $o=+1$ and $n=2$ for the left maximal puncture, and $o=-1$ and $n=-1$ for the right maximal puncture.
  The charge assignment to the bifundamental chiral multiplets is given by
    \bea
    q_{i}: R^{\half} t^{-\half} \gamma_{i} \alpha',~~~~
    \tilde{q}_{i}: R^{\half} t^{-\half} \beta_{i}^{-1} \alpha'^{-1}.
    \eea
  The $U(1)_{\alpha'}$ symmetry has mixed anomalies with $U(1)_{\beta_{i}}$ or $U(1)_{\gamma_{i}}$: 
  $\tr \beta_{i} \alpha^{2} = \tr \gamma_{i} \alpha^{2} = - kN^{2}$,
  which are the opposite signs to those in \eqref{betaanom1}.
    
  In what follows, we will refer to the building blocks of Figure \ref{fig:buildingblock1} 
  as $\CB^{++}_{+}$ and $\CB^{--}_{-}$ respectively.
  The reason for these signs becomes clear in a moment.

\subsection{Gauging}
  We now go back to the orbifolded SQCD.
  To get this theory from the building blocks of the section above, we gauge the diagonal $SU(N)^{k}$ group of $\widetilde{SU}(N)_{i}$ of 
  $\CB^{++}_{+}$ specified by the anomaly coefficients \eqref{betaanom1} and $SU(N)_{i}$ of $\CB^{--}_{-}$
  specified by the anomaly coefficients \eqref{betaanom2}.
  All the (diagonal) $U(1)$ global symmetries from the both sides are not broken in this gauging.
  For $\beta_{i}$ and $\gamma_{i}$ this is easy to see from the opposite signs 
  in \eqref{betaanom1} and \eqref{betaanom2}.
  The $U(1)_{t}$ is a diagonal part of $\beta_{i}$ or $\gamma_{i}$, so the anomalies are automatically canceled.
  For $R$, since $R(\lambda) = 1$ (where $\lambda$ is gaugino), 
  $\tr R SU(N)_{i}^{2} = N - \frac{1}{2} \frac{1}{2} 4N = 0$,
  where $SU(N)_{i}$ are gauge nodes.
  Therefore the residual global symmetry after gauging is $U(1)_{R} \times U(1)_{t} \times 
  \prod_{i} U(1)_{\beta_{i}} \times U(1)_{\gamma_{i}}
  \times SU(N)^{2k} \times U(1)_{\alpha} \times U(1)_{\alpha'}$.
  
  It is easy to count the number of the exactly marginal operators in the infrared fixed point
  by using the argument in \cite{Leigh:1995ep}.
  We have $k + k$ marginal couplings, which are gauge and quartic superpotential couplings.
  Then we assume that all the beta functions, written in terms of the anomalous dimensions, 
  vanish in the infrared.
  This gives $2k$ equations for $2k$ variables.
  However one of these equations is linearly dependent on the others since the quiver is cyclic.
  Therefore we have one exactly marginal coupling parametrizing the one-dimensional family of the solutions
  to the equations.
  
  The resulting theory is associated to a sphere with two maximal and two minimal punctures,
  where $SU(N)^{2k}$ symmetry corresponds to the former 
  and $U(1)_{\alpha}$ and $U(1)_{\alpha'}$ to the latter.
  The exactly marginal coupling possibly corresponds to the complex structure modulus of the four-punctured sphere.

\subsection{Dualities}
\begin{figure}[t]
	\centering 
	\includegraphics[width=7cm, bb=0 0 459 435]{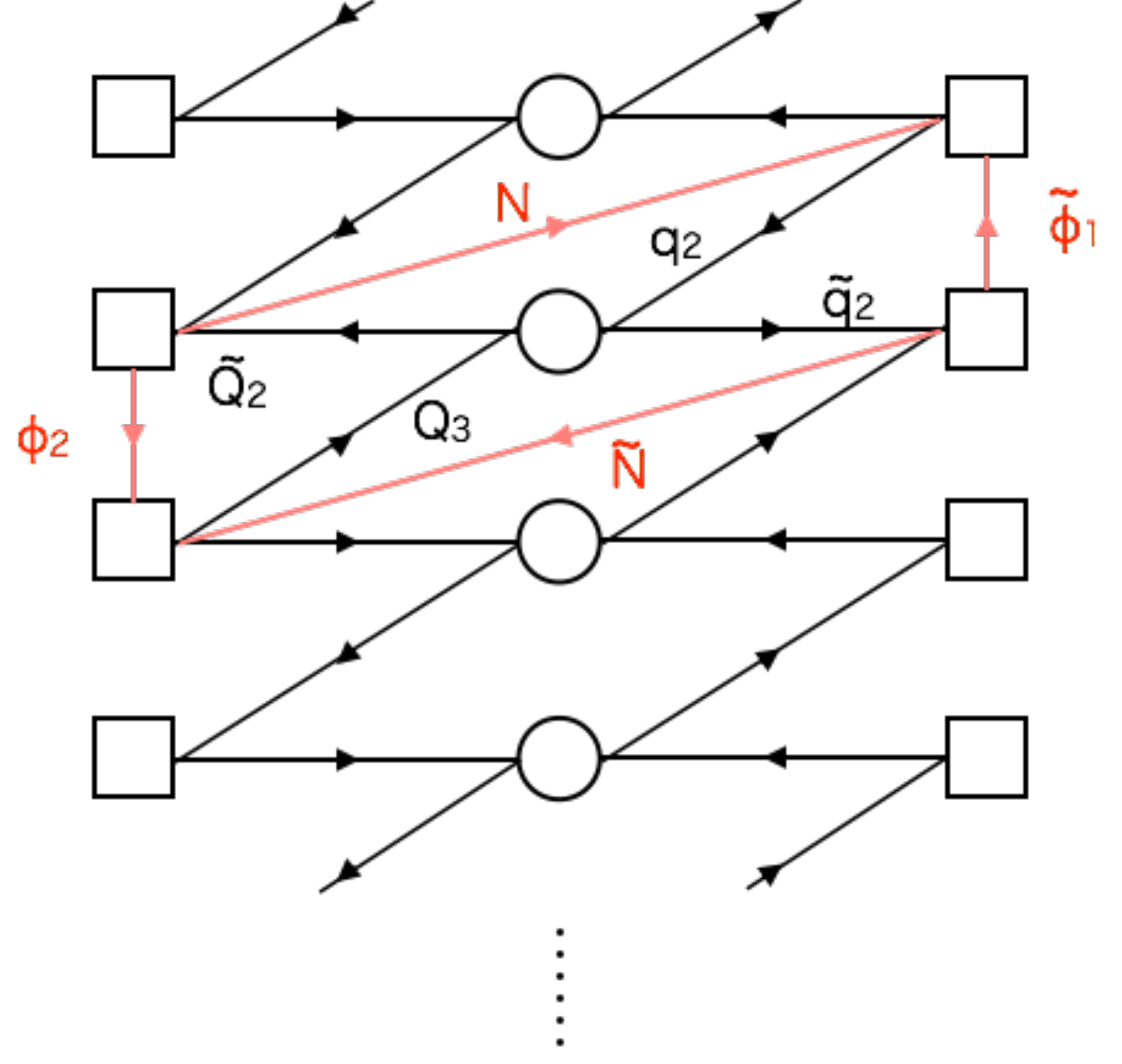}
	\caption{We take a Seiberg dual to the second gauge node in Figure \ref{fig:SQCD1}.
	         The arrows corresponding to $Q_{3}$, $\tilde{Q}_{2}$, $q_{2}$ and $\tilde{q}_{2}$ are flipped
	         and the chiral multiplets $\phi_{2}$, $\tilde{\phi}_{1}$, $N$ and $\tilde{N}$ are added.}
	\label{fig:SQCD2}
    \end{figure}

  Let us consider dualities of the orbifolded SQCD theory.
  Since all the gauge groups are effectively $SU(N)$ with $2N$ flavors, 
  Seiberg duality \cite{Seiberg:1994pq} does not change the rank of the gauge groups.
  Now let us first perform Seiberg duality on the second gauge node for example.
  The resulting theory is depicted in Figure \ref{fig:SQCD2}.
    The rules for computing Seiberg duality follow the usual rules of Urban Renewal \cite{Franco:2005rj}:
  The arrows of the chiral multiplets that are attached to the dualized gauge node are flipped,
  and the bifundamental ``mesons'' of the flavor symmetries are added.
  The charges of the new fields are as follows
    \bea
    &&
    \phi_{2}: R t \beta_{3} \gamma^{-1}_{2}, ~~~
    \tilde{\phi}_{1}: R t^{-1} \beta_{3}^{-1} \gamma_{2}, ~~~
    N, \tilde{N}: R,
    \nonumber \\
    & &
    \tilde{Q}_{2}: R^{\half} t^{-\half} \beta_{3}^{-1} \alpha'^{-1}, ~~~
    Q_{3}: R^{\half} t^{-\half} \gamma_{2} \alpha', ~~~
    q_{2}: R^{\half} t^{\half} \beta_{3} \alpha, ~~~
    \tilde{q}_{2}: R^{\half} t^{\half} \gamma_{2}^{-1} \alpha^{-1},
    \nonumber
    \eea
  and those of the other multiplets are unchanged.
  Because of the quartic superpotential in the original theory, we obtain the cubic superpotential
  on the dual theory
    \bea
    W
     =     \tr Q_{3} \tilde{Q}_{2} \phi_{2} + \tr q_{2} \tilde{q}_{2} \tilde{\phi}_{1}
         - \tr N q_{2} \tilde{Q}_{2} - \tr \tilde{N} Q_{3} \tilde{q}_{2}
         + \tr N \tilde{q}_{1} Q_{2} + \tr \tilde{N} \tilde{Q}_{3} q_{3},
    \eea
  plus quartic terms for the rhombi in Figure \ref{fig:SQCD2}.
  This in fact follows the usual graphical rule which was introduced in \cite{Hanany:1997tb} 
  to read off the superpotential: 
  closed paths in the quiver add to the superpotential with alternating signs, 
  depending on the orientation of the path.
     
\begin{figure}[t]
	\centering 
	\includegraphics[width=9cm, bb=0 0 580 435]{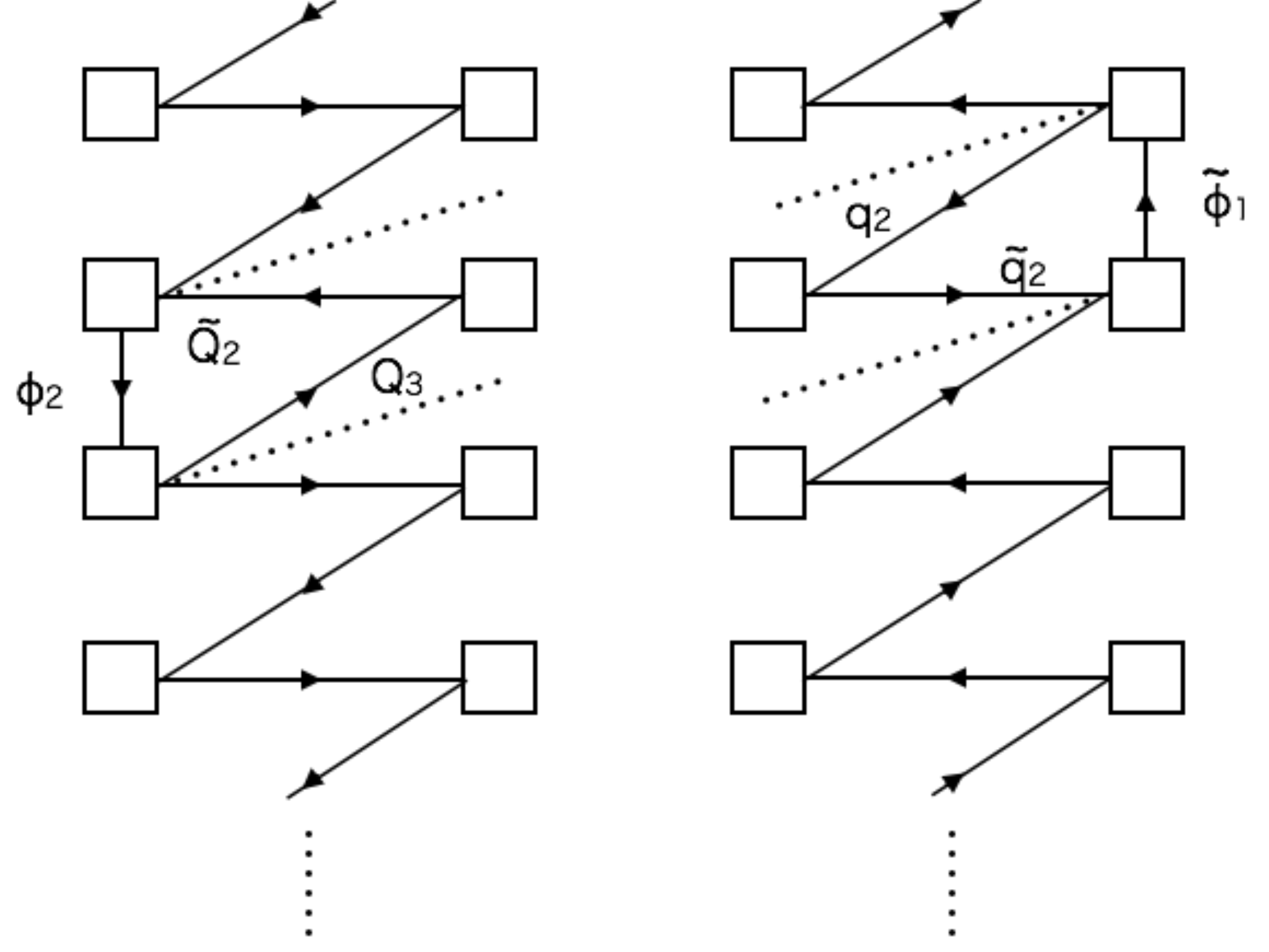}
	\caption{The building blocks of the theory in Figure \ref{fig:SQCD2}.}
	\label{fig:buildingblock3}
    \end{figure}

  Before going to another dual description, let us consider the new building blocks which are seen in the quiver 
  of Figure \ref{fig:SQCD2}.
  By decoupling the $SU(N)$ gauge nodes, we get a pair of building blocks as in Figure \ref{fig:buildingblock3}.
  The dashed lines correspond to $N$ and $\tilde{N}$ multiplets.
  Due to their charges they do not contribute to any anomalies. 
  Let us focus on the left building block.
  One can easily check that the the anomalies of $U(1)_{\beta_{i}}$ (or $U(1)_{\gamma_{i}}$)
  and $SU(N)_{i}$ are unchanged.
  However the anomalies of $\widetilde{SU}(N)_{i}$ are as follows:
    \bea
    \tr \beta_{i+1} \widetilde{SU}(N)_{i}^{2}
     =   - \tr \gamma_{i} \widetilde{SU}(N)_{i}^{2}
     =     \left\{ \begin{array}{ll}
           N/2 & (i\neq 2) \\ 
         - N/2 & (i=2) \\
           \end{array} \right.
           \label{betaanom3}
    \eea 
  The $\tr R SU(N)_{i}^{2}$ and $\tr R \widetilde{SU}(N)_{i}^2$ are unchanged, but
    \bea
    \tr T
     =     (k-1)N^{2}, ~~~
    \tr T^{3}
     =     \frac{(k+2)N^{2}}{4}.
    \eea
  The $SU(N)^{k}$ and $\widetilde{SU}(N)^{k}$ symmetries are associated to two maximal punctures
  as in the previous case, and we assign a color $n$ and an orientation $o$ to the puncture.
  However the anomalies \eqref{betaanom3} indicate that 
  this assignment is not enough to classify all possible building blocks,
  and instead there seems to be signs that are attached to each $SU(N)$ symmetry.
  Therefore we label a maximal puncture by $n$, $o$, and $\sigma_{i} (=\pm1)$
  where $i=0,\ldots,k-1$.
  
  There is also a subtlety in the $U(1)_{\alpha}$ and $U(1)_{\alpha'}$ symmetries coming from the minimal punctures.
  The anomalies of these with $\beta_{i}$ and $\gamma_{i}$ are given by 
    \bea
    \tr \beta_{i+1} \alpha^{2}
     =   - \tr \gamma_{i} \alpha^{2}
     =     \left\{ \begin{array}{ll}
           N^{2} & (i\neq 2) \\ 
           0 & (i=2) \\
           \end{array} \right., ~~~
    \tr \beta_{i+1} \alpha'^{2}
     =   - \tr \gamma_{i} \alpha'^{2}
     =     \left\{ \begin{array}{ll}
           0 & (i\neq 2) \\ 
         - N^{2} & (i=2) \\
           \end{array} \right..
    \eea
  Thus, this building block depends not only on the minimal puncture with $U(1)_{\alpha}$ 
  but also on the puncture with $U(1)_{\alpha'}$.
  To label this we introduce another collection of signs $\sigma_{i}^{(min)}$. 
  For the current example the sign assignment is $(+,+,-,+,\ldots,+)$.
  In summary, the left building block in Figure \ref{fig:buildingblock3} is associated to a sphere with
  a left maximal puncture that is labeled by $n=1$, $o=1$ and $\sigma_{i}=(+,\ldots,+)$, 
  a right maximal puncture with $n=0$, $o=-1$
  and $\sigma_{i} = (+,+,-,+,\ldots,+)$, and a minimal puncture with $\sigma_{i}^{(min)} = (+,+,-,+,\ldots,+)$
    
\begin{figure}[t]
	\centering 
	\includegraphics[width=5.8cm, bb=0 0 381 435]{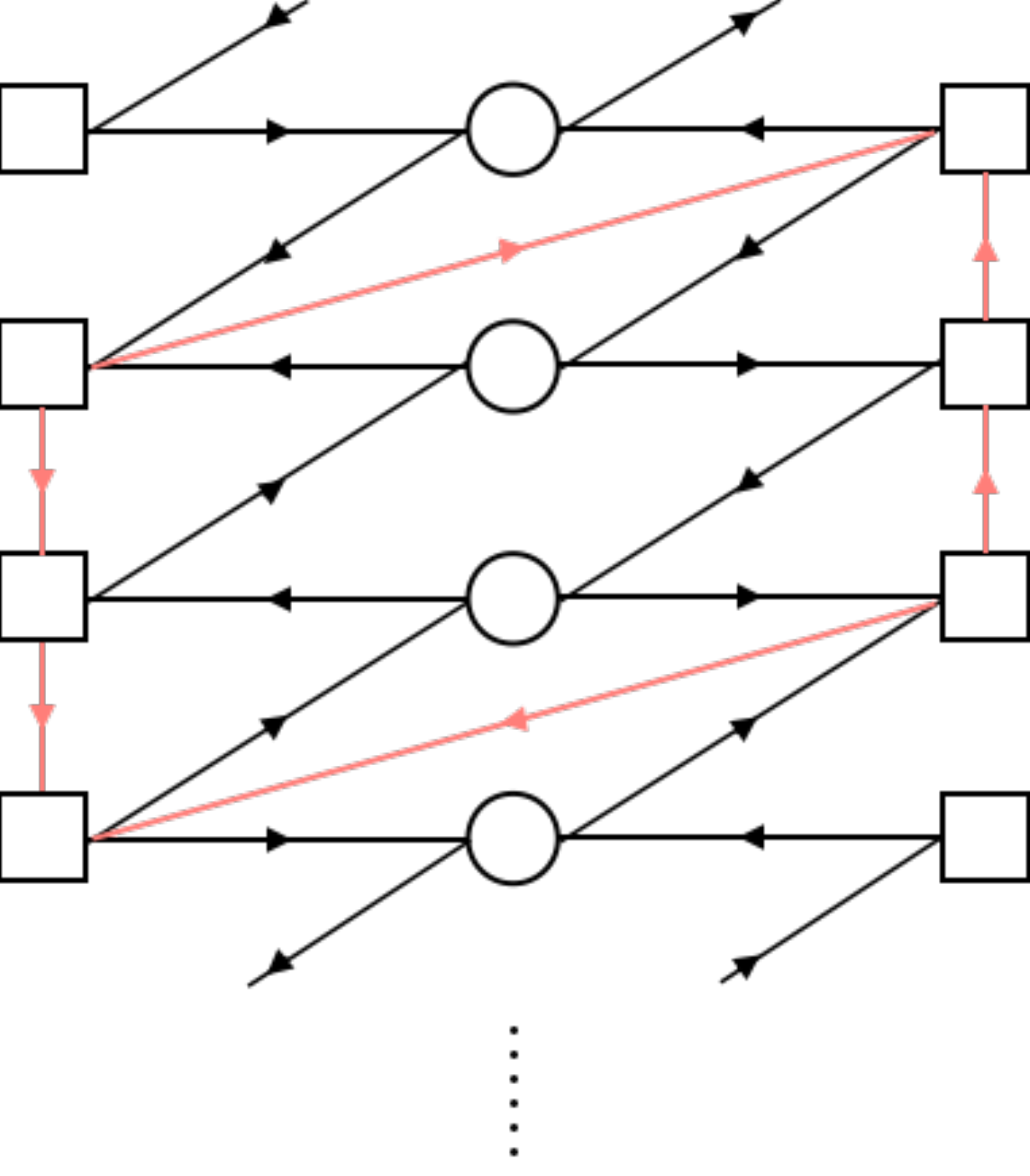}
	\caption{The Seiberg dual to the third node in Figure \ref{fig:SQCD2}.}
	\label{fig:SQCD3}
    \end{figure}

  Now let us take another Seiberg dual to the quiver in Figure \ref{fig:SQCD2}.
  If we dualize at the same node (namely at the $SU(N)_{2}$ gauge group again) we get back to the original theory.
  We can get another quiver by taking a dual on another node.
  For instance, Seiberg duality on the third node gives the quiver depicted in Figure \ref{fig:SQCD3}.
  Here the $\tilde{N}$ multiplet becomes massive and is integrated out, and
  other chiral multiplets appear \cite{Franco:2005rj}.
  
  The left building block of this quiver is specified by
  a left maximal puncture with $n=1$, $o=1$ and $\sigma_{i}=(+,\ldots,+)$, 
  a right maximal puncture with $n=0$, $o=-1$
  and $\sigma_{i} = (+,+,-,-,\ldots,+)$, and a minimal puncture with $\sigma_{i}^{(min)} = (+,+,-,-,\ldots,+)$.
  
\begin{figure}[t]
	\centering 
	\includegraphics[width=14cm, bb=0 0 1008 445]{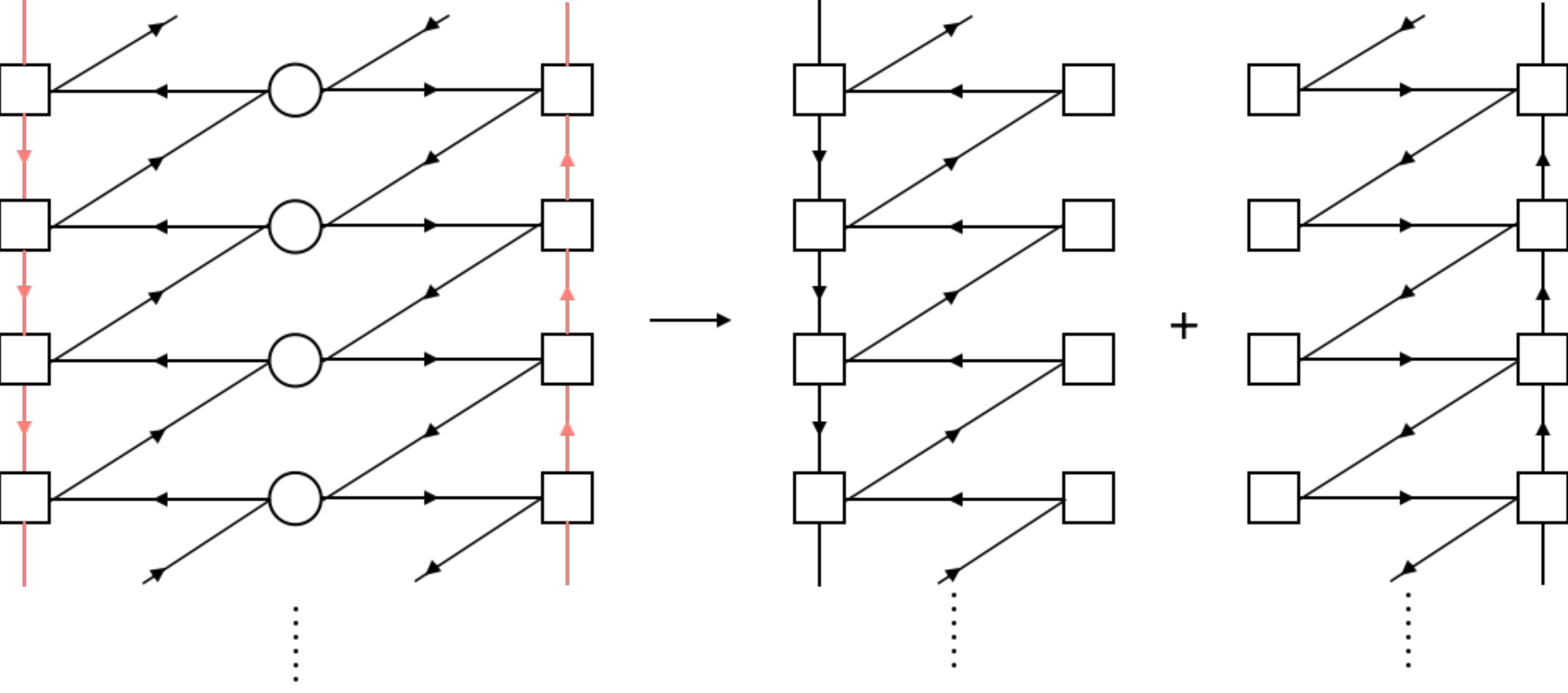}
	\caption{Left: A dual description where all the gauge groups are dualized.
	         Right: Decomposition to two building blocks.
	         Both building blocks have $\sigma_{i} = 1$ and $\tilde{\sigma}_{i} = -1$,
	         but $\sigma_{i}^{(min)} = -1$ for the left and $\sigma_{i}^{(min)} = +1$ for the right.}
	\label{fig:SQCD4}
    \end{figure}

  Finally the extreme case is to take Seiberg dualities at all the gauge nodes.
  This gives the quiver in Figure \ref{fig:SQCD4}.
  For all the flavor $SU(N)$ groups bifundamental chiral multiplets, $\phi_{i}$, are attached.
  One building block of this quiver is depicted in the middle of Figure \ref{fig:SQCD4}, and
  the charges of the chiral multiplets are given by
    \bea
    Q_{i}: R^{\half} t^{-\half} \gamma_{i-1} \alpha', ~~~
    \tilde{Q}_{i}: R^{\half} t^{-\half} \beta_{i+1}^{-1} \alpha'^{-1}, ~~~
    \phi_{i}: R t \beta_{i+1} \gamma^{-1}_{i}.
    \eea
  Thus the anomaly coefficients are 
  $\tr \beta_{i} SU(N)_{i}^{2} = N/2$ and $\tr \beta_{i+1} \widetilde{SU}(N)_{i}^{2} = -N/2$.
  This is specified by a left maximal puncture with $n=1$, $o=1$ and $\sigma_{i}=(+,\ldots,+)$, 
   a right maximal puncture with $n=0$, $o=-1$ and $\sigma_{i} = (-,\ldots,-)$, 
   and $\sigma_{i}^{(min)} = (-,\ldots,-)$.
  We also have anomalies $\tr T = 0$ and $\tr T^{3} = 3kN^{3}/4$.
  Note also that this building block does not depend on $U(1)_{\alpha}$.
  
  Actually this quiver can be obtained by the orbifold projection of the Seiberg dual theory of
  the $\CN=1$ $SU(N)$ SQCD which we saw at the beginning of this Section.
  The dual theory is again an $SU(N)$ SQCD with $N+N$ flavors and two chiral multiplets
  transforming in the adjoint representations of $SU(N)^{2}$ flavor symmetries respectively,
  and with a quartic coupling (see \cite{Gadde:2013fma} for the dual description).
  The only difference from the original theory is the addition of the adjoint chiral multiplets.
  Thus we need to consider the projection of these, which have $U(1)_{\CF}$ charge $+1$ and $-1$.
  It is easy to see that the projection give bifundamental chiral multiplets 
  in $(\bf{N}_{i}, \bar{\bf{N}}_{i+1})$ and $(\bar{\tilde{\bf{N}}}_{i}, \tilde{\bf{N}}_{i+1})$ 
  of the flavor $SU(N)$ symmetries,
  which are indeed the bifundamentals (the red vertical ones) in Figure \ref{fig:SQCD4}.
  
  We have $2^{k}$ dual descriptions including the original quiver.
  These correspond to the possible choices of $\sigma_{i}$ of the maximal puncture which is glued.
  (The $\tilde{\sigma}_{i}$'s of the other glued puncture from the other building block is automatically fixed 
  to be $\tilde{\sigma}_{i} = - \sigma_{i}$ due to anomaly cancelation.)
  Among them the two descriptions in Figures \ref{fig:SQCD1} and \ref{fig:SQCD4} have 
  a simple Type IIA brane configuration. 
  They are obtained as orbifold projections of a brane system and its dual by the exchange of NS5-branes.
  Thus these two cases fall into a special subclass of the chiral theories - those with simple Type IIA counterpart.

\section{Chiral theories of class $\CS$}
\label{sec:class}

\subsection{Classification of building blocks}
\label{subsec:bb}
  In the previous section we find the classification parameters of different building blocks by performing Seiberg dualities on orbifolded SQCD. Let us summarize the current picture. 
  A building block with $SU(N)^{k} \times \widetilde{SU}(N)^{k}$ symmetries 
  can be classified by two maximal punctures with $\sigma_{i}$, $\tilde{\sigma}_{i}$
  and $n$ and $o$, and one minimal puncture with $\sigma_{i}^{(min)}$. The existence of the theory for an arbitrary choice of signs $\sigma_{i}$ and $\tilde{\sigma}_{i}$ is not clear, but we will work under the assumption that such a theory exists unless extra conditions arise. Let us turn to a detailed analysis.

\paragraph{Classification}
  We classify the basic building blocks associated to two maximal and one minimal punctures as follows:
    \begin{itemize}
    \item to each maximal puncture we assign the color $n$ ($n=0,1,\ldots,k-1$), the orientation $o$ ($o=+1$ or $-1$)
    and flavor symmetries $\prod_{i} SU(N)_{i}$ with the signs $\sigma_{i}$ ($i=0,\ldots,k-1$),
    \item to each minimal puncture we assign the signs $\sigma^{(mim)}_{i}$,
    and a symmetry $U(1)_{\alpha}$,
    \end{itemize}
  We always choose the opposite orientations for two maximal punctures in what follows. 
  In addition to these, we can introduce discrete curvatures of $U(1)_{t}$, 
  $U(1)_{\beta_{i}}$ and $U(1)_{\gamma_{i}}$.
  This issue is discussed in Section \ref{sec:higgs}.  
    
  In the examples studied above, we have
    \begin{itemize}
    \item building block in the left of Figure \ref{fig:buildingblock1}, 
    $\sigma_{i}=\tilde{\sigma}_{i} = \sigma_{i}^{(mim)} =+$.
    \item building block in the right of Figure \ref{fig:buildingblock1}, 
    $\sigma_{i}=\tilde{\sigma}_{i} = \sigma_{i}^{(mim)}  =-$
    \item building block in the left of Figure \ref{fig:buildingblock3}, 
    $\sigma_{i}= +$ and $\tilde{\sigma}_{i} = \sigma_{i}^{(mim)} = (+,+,-,\ldots,+)$.
    \item building block in the right of Figure \ref{fig:buildingblock3}, 
    $\tilde{\sigma}_{i}= -$ and $\sigma_{i} = \sigma_{i}^{(mim)} = (-,-,+,\ldots,-)$.
    \item building block in the center of Figure \ref{fig:SQCD4}, 
    $\sigma_{i}= +$ and $\tilde{\sigma}_{i} = \sigma_{i}^{(mim)} = -$.
    \end{itemize}
  
  It is also easy to flip all the signs of the building blocks by charge conjugation of chiral multiplets.
  Graphically this corresponds to keeping all the matter contents but flipping their arrows.

\begin{figure}[t]
	\centering 
	\includegraphics[width=12cm, bb=0 0 1136 571]{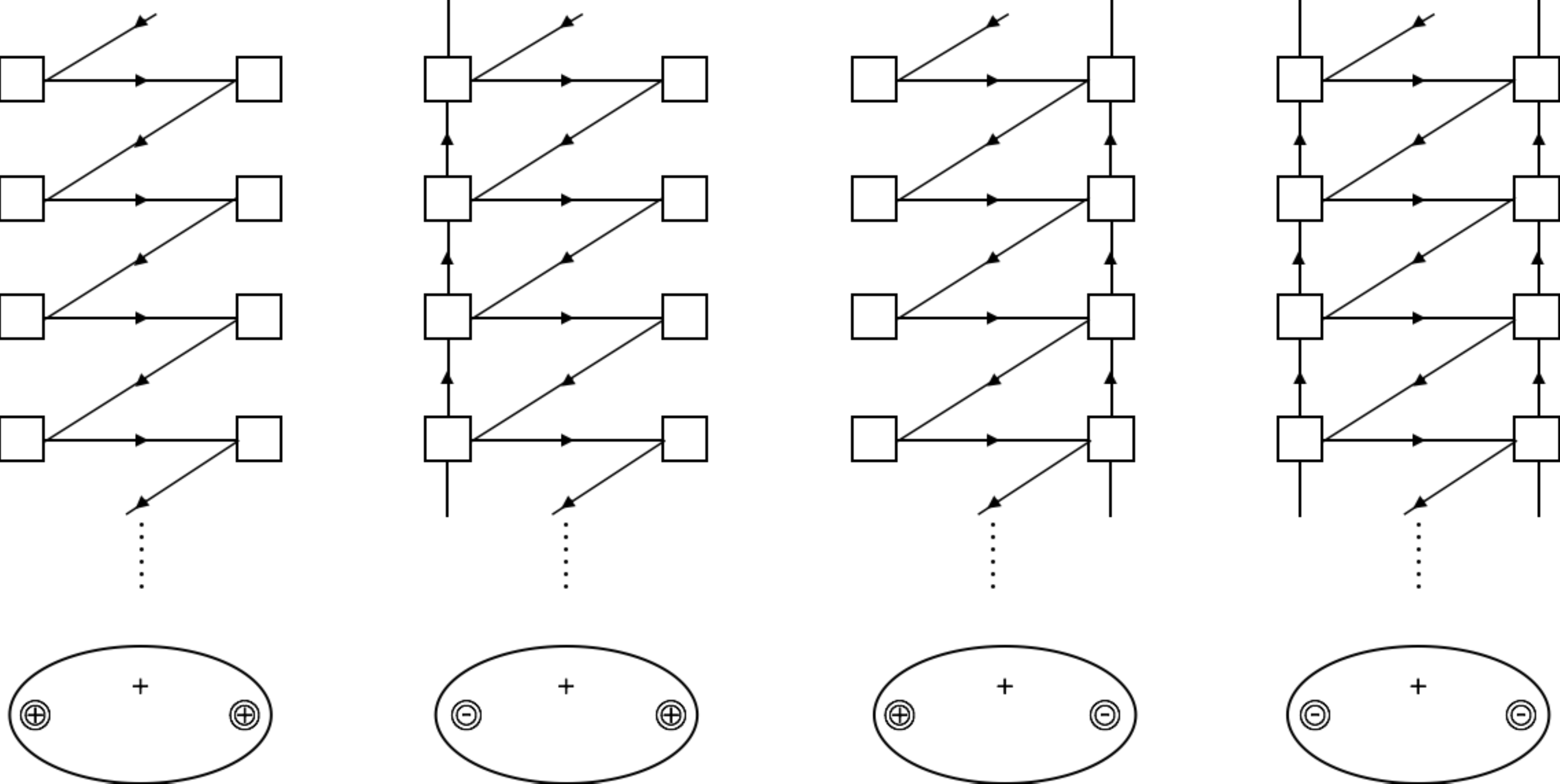}
	\caption{Four building blocks and associated pairs-of-pants with $\sigma^{(min)} = +1$.
	         The signs in the double circle correspond to those of the maximal punctures.
	         The signs without circle correspond to those of the minimal punctures.
	         These theories are referred to as $\CB^{++}_{+}$, $\CB^{-+}_{+}$, $\CB^{+-}_{+}$, $\CB^{--}_{+}$.}
	\label{fig:buildingblock5}
    \end{figure}

\paragraph{A special subset}
  A subset of the above classification is the building blocks where 
  each puncture has one definite sign, namely $\sigma_{i}$ are the same for all $i$.
  As stated in the end of the previous section, this restriction corresponds to the class of theories 
  which have Type IIA counterparts.
  We will see this in subsequent sections, 
  and use a shorthand notation for these signs $\sigma \equiv \sigma_{i}$.
  For $\sigma^{(min)}=+1$, we have four different building blocks shown in Figure \ref{fig:buildingblock5}.
  We can see that the sign of the maximal puncture corresponds to the presence/absence
  of the bifundamental multiplets, $\phi_{i}$, of $SU(N)_{i} \times SU(N)_{i+1}$ 
  and the cubic superpotential term for each triangle.
  
  For later convenience, let us denote these four building blocks as
    \bea
    \CB^{++}_{+}, ~~~
    \CB^{-+}_{+}, ~~~
    \CB^{+-}_{+}, ~~~
    \CB^{--}_{+},
    \eea
  where the subscript denotes the sign of the minimal puncture 
  and the superscripts denote the signs of the two maximal punctures.
  The $\CB_{+}^{++}$ has already appeared in Section \ref{sec:SQCD}.
  The colors of the left and the right maximal punctures are $n=1$ and $n=0$ respectively.
  $\CB_{+}^{-+}$ is described by bifundamental chiral multiplets 
  $Q_{i}: R^{\half} t^{\half} \beta_{i} \alpha$ in $(\bar{\bf{N}}_{i-1}, \bf{\tilde{N}}_{i-1})$,
  $\tilde{Q}_{i}: R^{\half} t^{\half} \gamma_{i}^{-1} \alpha^{-1}$ in $(\bf{N}_{i-1}, \bar{\bf{\tilde{N}}}_{i})$,
  and $\phi_{i}: R t^{-1} \beta_{i+2}^{-1} \gamma_{i+1}$ in $(\bar{\bf{N}}_{i}, \bf{N}_{i+1})$.
  The colors of the left and the right maximal punctures are $n=3$ and $n=0$ respectively.
  The other two building blocks can be described in a similar way:
  the colors of the two maximal punctures are $n=1$ and $n=-2$ for $\CB_{+}^{+-}$,
  and $n=3$ and $n=-2$ for $\CB_{+}^{--}$.
  Of course the colors are shifted in the same amount 
  by changing the definition of $U(1)_{\beta}$ while fixing $U(1)_{\gamma}$.
  
  There exist ``meson operators'', $M_{i}^{~i+1}$, 
  which are bifundamental of $SU(N)_{i} \times SU(N)_{i+1}$ flavor symmetry.
  In the building block (the leftmost one in Figure \ref{fig:buildingblock5}), 
  the mesons of the left and right flavor symmetries are $M_{i}^{~i+1} = \tilde{Q}_{i} Q_{i+1}$ and
  $\tilde{M}_{i}^{~i+1} = Q_{i+1} \tilde{Q}_{i+1}$ respectively.
  In the second building block in Figure \ref{fig:buildingblock5}, 
  the mesons of the left and right flavor symmetries are $M_{i}^{~i+1} = \phi_{i}$ and 
  $\tilde{M}_{i}^{~i+1} = Q_{i+1} \tilde{Q}_{i+1}$ respectively, and so on.
  
  The charge conjugation of these building blocks are denoted in a similar way as
  $\CB^{--}_{-}$, $\CB^{+-}_{-}$, $\CB^{-+}_{-}$ and $\CB^{++}_{-}$.
  To abbreviate these, we will also use the notation $\CB^{\sigma \tilde{\sigma}}_{\sigma^{(b)}}$.

\paragraph{Other building blocks}
  The above building blocks are actually free in the infrared.
  (This is checked in Section \ref{sec:anomaly}.)
  One could wonder the possibility to have building blocks which are interacting SCFTs in the infrared.
  Indeed, the case with three maximal punctures corresponds 
  to an orbifolded version of the $T_{N}$ theory \cite{Gaiotto:2009we}, 
  which might be an interacting SCFT in the orbifold case as well.
  Furthermore we can have a puncture associated with a flavor symmetry which is a subgroup of $SU(N)^{k}$
  obtained by a particular Higgsing of the maximal puncture.
  Although it would be interesting to pursue this direction, we will not discuss these cases.
  Instead, we will see in Section \ref{sec:higgs} that even a pair-of-pants 
  with two maximal and one minimal punctures gives rise to a nontrivial SCFT in the infrared.
  Before going to the details, let us study the building blocks from the M-/string theory viewpoint.

\subsection{M5-brane compactification}
  We propose that some of the theories above are obtained from a compactification of the M5-branes.
  Since these are obtained by taking the orbifold of $\CN=1$ class $\CS$ theories 
  \cite{Bah:2012dg,Beem:2012yn,Gadde:2013fma,Bah:2013aha,
  Xie:2013gma,Agarwal:2013uga,Agarwal:2014rua,McGrane:2014pma}
  we first review the latter.
  (See also \cite{Maruyoshi:2013hja,Bonelli:2013pva,Xie:2013rsa,Yonekura:2013mya} for non-conformal cases)
  Let us consider three-(complex) dimensional manifold $\CL_{1} \oplus \CL_{2}$
  over a Riemann surface $C_{g,n}$
  where $\CL_{1,2}$ are line bundles, $g$ is the genus and $n$ is the number of punctures.
  We impose a condition $\CL_{1} \otimes \CL_{2} = T^{*} C_{g,n}$ 
  which corresponds to $p + q = 2g - 2 + n$, where $p$ and $q$ are the degrees of $\CL_{1}$ and $\CL_{2}$ respectively.
  Then we put $N$ M5-branes on $\mathbb{R}^{1,3} \times C_{g,n}$.
  Note that the fibre directions of the line bundles are transverse to the worldvolume of the M5-branes.
  The theory on $\mathbb{R}^{1,3}$ preserves $\CN=1$ supersymmetry, and is classified by $C_{g,n}$,
  to which the M5-branes are compactified on.
  The classification is as follows: to each puncture 
  we associate a flavor symmetry related to a Young diagram and an additional label $\sigma$, 
  which chooses a fibre direction of the singularity;
  to each pair-of-pants we associate a sign $\sigma^{(b)}$.
  The degrees $p$ and $q$ correspond to the numbers of pairs-of-pants 
  with $\sigma^{(b)}=+$ and $\sigma^{(b)}=-$, respectively.
  (In \cite{Bah:2012dg,Agarwal:2015vla}, more exotic pair-of-pants with $p$, $q$ arbitrary was considered, 
  we will not consider this case here.)
  There are two isometries from the line bundles $U(1)_{\pm}$.
  The diagonal combination $U(1)_{+} + U(1)_{-}$ is used to cancel the holonomy of the Riemann surface
  and gives rise to $U(1)_{R}$, 
  and the other combination $U(1)_{\CF}$ is an additional global symmetry of the class $\CS$ theories. 
  
  Now let the fibre coordinates of $\CL_{1}$ and $\CL_{2}$ be $v$ and $w$ respectively.
  We consider the orbifold action as
    \bea
    v \rightarrow e^{2 \pi i/k} v, ~~~
    w \rightarrow e^{-2\pi i/k} w.
    \eea
  This orbifold still preserves $\CN=1$ supersymmetry in four dimensions.
  The compactified Riemann surface is still the same, but the information attached to the Riemann surface
  and the punctures are affected since they come from the nontrivial fibers.
  As we will see shortly the signs assigned to punctures and pairs-of-pants descend to 
  the signs seen in the previous subsection.
  
  Note that in the classification in Section \ref{subsec:bb} the sign $\sigma^{(b)}$
  associated to the pair-of-pants is not included.
  The reason might be that for the building blocks which have known Lagrangian descriptions
  the signs of the minimal puncture and the pair-of-pants are equal.
  This is true in the $\CN=1$ class $\CS$ theories \cite{Gadde:2013fma}, 
  and it is natural to suppose this is the case even in the class of chiral theories.
  Although adding the signs associated to pairs-of-pants is somehow redundant for the building blocks
  which we are considering, 
  we denote this below as $\sigma^{(b)}$ to keep it as general as possible.
  Again the numbers of the pairs-of-pants with $\sigma^{(b)} = +$ and $-$ are equal to $p$ and $q$
  discussed above respectively.
  
  An alternative viewpoint of the class of chiral theories is as a compactification of the
  six-dimensional $(1,0)$ theory on a Riemann surface.
  This six-dimensional theory is obtained 
  on a worldvolume of $N$ M5-branes on the $\IC^{2}/\IZ_{k}$ orbifold singularity.
  The tensor branch of this theory can be seen by the reduction to Type IIA string theory
  with $k$ D6-branes and $N$ NS5-branes, where D6-branes suspended between neighboring NS5-branes
  and streatched to $\pm \infty$ along the $x^{9}$-direction. 
  The six-dimensional theory consists of bifundamental hypermultiplets
  of $SU(k)_{h} \times SU(k)_{h+1}$ ($h=0,1,\ldots,N$), $SU(k)_{h}$ vector multiplets ($h=1,2,\ldots,N-1$),
  and $N-1$ tensor multiplets.
  The global symmetry is $SU(2)_{R} \times U(1)_{t} \times SU(k)_{0} \times SU(k)_{N}$
  where the $U(1)_{t}$ is the non-anomalous combination of the $U(1)$'s of bifundamental hypermultiplets.
  
  When compactified to four dimensions these symmetries descend to the intrinsic symmetry
  $U(1)_{R} \times U(1)_{t} \times (\prod_{i} U(1)_{\beta_{i}})/U(1) \times (\prod_{i} U(1)_{\gamma_{i}})/U(1)$ 
  which we already saw above.
  In principle one can introduce the curvature of the latter $U(1)$'s.
  The curvature of $U(1)_{t}$ is related to the colors and the signs associated to the punctures,
  which will be discussed in the next subsection.

\subsection{Branes and $U(1)_{t}$ curvature}
\label{subsec:brane}
\begin{figure}[t]
	\centering 
	\includegraphics[width=13cm, bb=0 0 876 302]{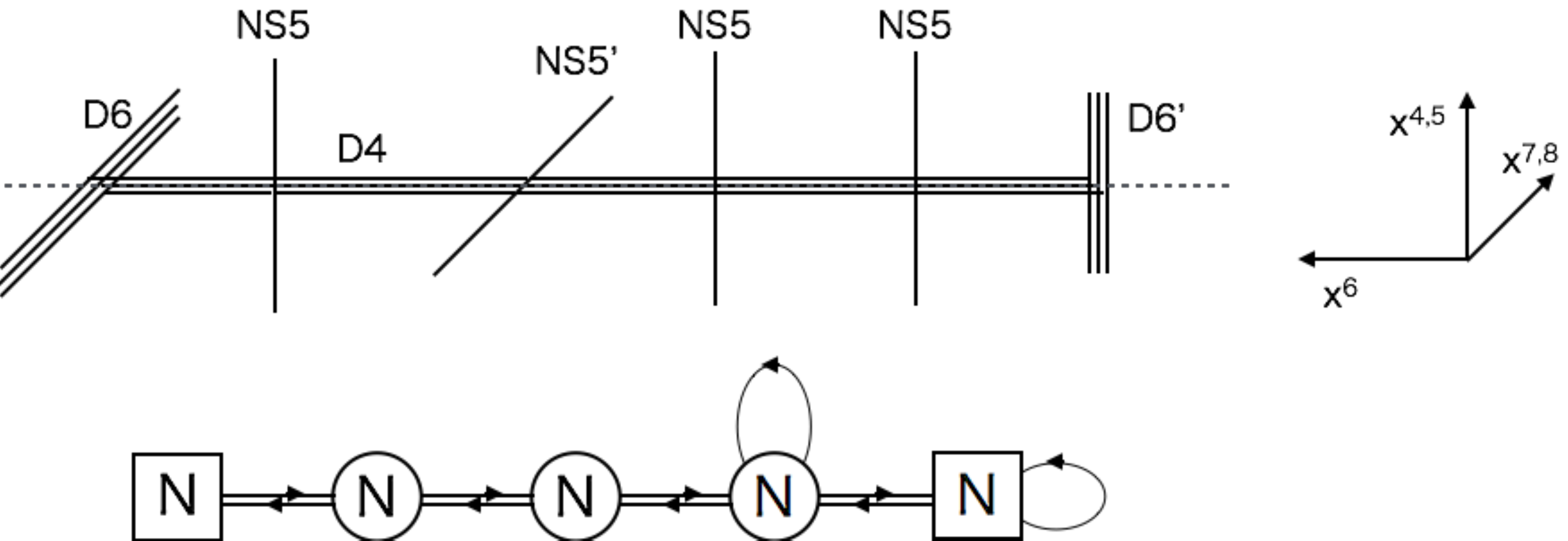}
	\caption{A Type IIA brane configuration of a linear quiver theory.
	         The D4 branes suspended between NS5-branes of the same type give an additional chiral multiplet
	         in the adjoint representation of the gauge group.
	         Also the D4-branes suspended between NS5 and D6'-branes give an additional chiral multiplet
	         in the adjoint representation of the flavor group.}
	\label{fig:branelinear}
    \end{figure}

  To see the effect of the $U(1)_{t}$ curvature, 
  let us go back to the M-theory configuration in the beginning of the previous subsection,
  and reduce it to Type IIA by compactifying not on the circle in $\IC^{2}/\IZ_{k}$, 
  but on the $S^{1}$ in the Riemann surface.
  We restrict the Riemann surface to a cylinder or a torus, and then 
  get a Type IIA brane system \cite{Witten:1997sc,Elitzur:1997hc} with the orbifold action on the $v$ and $w$ planes.
  In addition to D4-, NS5- and NS5$'$-branes already seen in the beginning of Section \ref{sec:SQCD} 
  we add D6- and D6'-branes which occupy $x^{i}$, $i=0,1,2,3,7,8,9$ and $i=0,1,2,3,4,5,9$ directions respectively.
  An example of the brane configuration associated to a cylinder is depicted in Figure \ref{fig:branelinear},
  where the orbifold singularity is at the origin of $x^{4,5}$ and $x^{7,8}$.
  The NS5- and NS5$'$-branes correspond to the minimal punctures on the cylinder
  whose signs are related to two types of NS5-branes, $\sigma^{(min)}=+1$ for NS5 and $-1$ for NS5$'$.
  The two boundaries of the cylinder are associated to two maximal punctures
  whose signs are related to two types of D6-branes, $\sigma^{(max)} = +1$ for D6 and $-1$ for D6$'$. 
  The toric case where the $x_{6}$-direction is compactified can be considered in the similar way.
  
  The orbifold projection of this kind of four-dimensional theory has been already studied in Section \ref{sec:SQCD}.
  The additional chiral multiplets give a set of bifundamental chiral multiplets in the vertical direction.
  
  Now we can go a bit further from this brane configuration by T-dualizing in the $x^{5}$ direction.
  To simplify the problem let us focus only on the part corresponding to the simplest building block
  as in Figure \ref{fig:branebuildingblock}:
    \begin{figure}[t]
	\centering 
	\includegraphics[width=13cm, bb=0 0 1019 329]{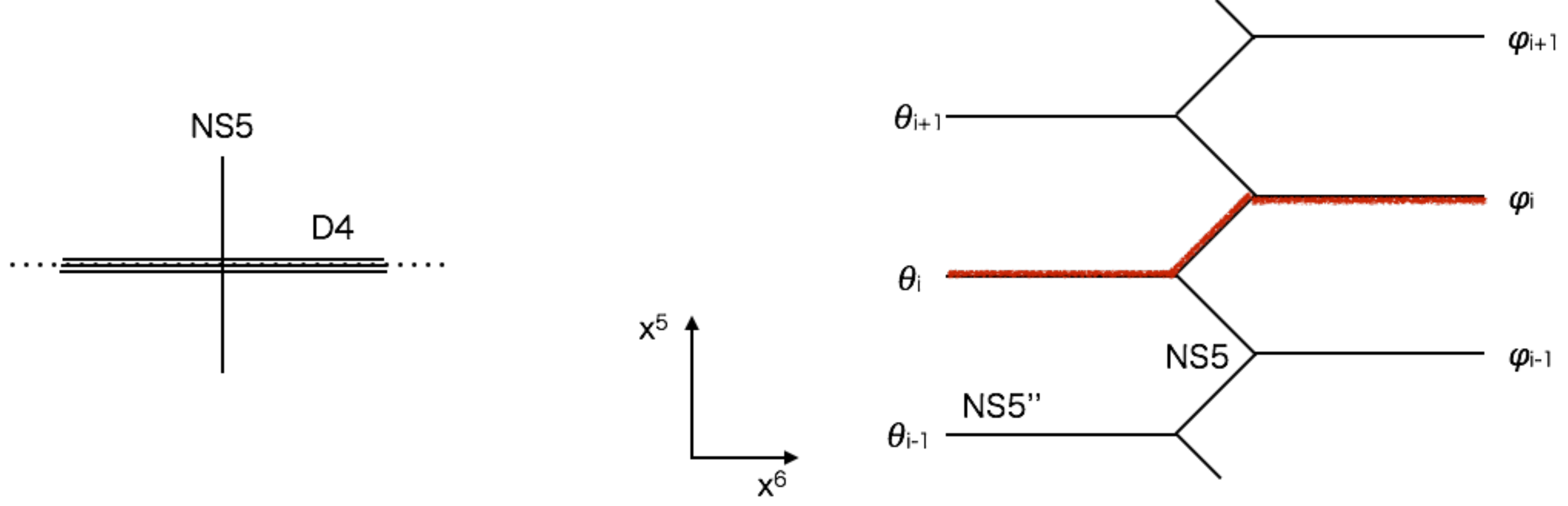}
	\caption{The brane configurations of the building block $\CB_{+}^{++}$ 
	in Type IIA (left) and in Type IIB (right).}
	\label{fig:branebuildingblock}
    \end{figure}
  one NS5-brane is attached by $N$ D4-branes from the both sides 
  (we took the position of the D6-branes to be $\pm \infty$). 
  In presence of the orbifold this is mapped to 
  D5-, NS5-, NS5$''$-brane system in Type IIB.
  The D5-, NS5-, and NS5$''$-branes are occupying $i=0,1,2,3,5,6$, $i=0,1,2,3,4,5$, and $i=0,1,2,3,6,9$ directions
  respectively.
  (The NS5$''$-brane comes from T duality on the orbifold.)
  In the Figure \ref{fig:branebuildingblock} drawn on $x^{5,6}$ plane,
  there are $N$ D5-branes in each region surrounded by NS5- and NS5$''$-branes (and infinity). 
  This is a typical brane tiling configuration \cite{Hanany:2005ve,Franco:2005rj} 
  where each tile corresponds to an $SU(N)$ flavor symmetry 
  (as the area of the tile is proportional to the inverse gauge coupling).
  Note that the $x^{5}$-direction is compactified and the rotation symmetry is interpreted as $U(1)_{t}$.
  An NS5-brane between D5-branes corresponds to a chiral bifundamental multiplet.
  In this way we recover the building block $\CB_{+}^{++}$.
  
  The $\pm$ infinities in $x^{6}$ directions correspond to the two maximal punctures.
  Let us inspect this in more detail.
  The $SU(N)^{k}$ flavor symmetry for each side of the NS5-brane is associated, of course, to a maximal puncture.
  Furthermore we have seen that the color $n$ and the sign $\sigma$ should be specified.
  To see this, let us parametrize the positions of the NS5$''$-branes at the $-$ and $+$ infinities
  by the angles $\theta_{i}$ and $\varphi_{i}$ respectively.
  Let the NS5$''$-branes be equally spaced, namely separated by a $2\pi/k$ angle.
  Then the angles are given by $\theta_{i} = \left( i + \frac{1 - n_{L}}{2} \right) (\frac{2\pi}{k})$ and 
  $\varphi_{i}= \left( i + \frac{1 - n_{R}}{2} \right) (\frac{2\pi}{k})$, 
  where $n_{L}$ ($n_{R}$) are the color of the left (right) maximal puncture respectively.
  The building block $\CB_{+}^{++}$ with $n_{L}=1$ and $n_{R}=0$ corresponds to angles
  $\theta_{i} = i (\frac{2\pi}{k})$ and $\varphi_{i} = (i+\half) (\frac{2\pi}{k})$.
  
  Since the rotation symmetry in the $x^{5}$-direction is $U(1)_{t}$, the $1/2$ overall shift 
  between $\theta_{i}$ and $\varphi_{i}$ is interpreted as
  the introduction of a $U(1)_{t}$ curvature.
  This shift is due to the NS5-brane which is associated to the minimal puncture.
  Therefore we interpret of the sign $\sigma^{(min)}$ of the minimal puncture 
  exactly as the introduction of $U(1)_{t}$.
  Interestingly the $U(1)_{t}$ curvature is traced by the zig-zag path \cite{Hanany:2005ss} 
  on the brane tiling picture,
  as drawn as a red line in Figure \ref{fig:branebuildingblock}.
  This is as it should be, because generally a zig-zag path corresponds to a $U(1)$ symmetry.
  The symmetry associated to $i$-th path is $U(1)_{\beta_{i}}$, which connect $\theta_{i}$ and $\varphi_{i}$.
  Thus this leads to an overall shift of all $U(1)_{\beta_{i}}$ which is $U(1)_{t}$.
  
  In general the maximal punctures with $\sigma_{L}$ for the left and $\sigma_{R}$ for the right,
  have $x^{5}$ angles which are given by
    \bea
    \theta_{i}
     =     \left( i - \frac{n_{L} + \sigma_{L}}{2} +1 \right) \frac{2\pi}{k}, ~~~
    \varphi_{i}
     =     \left( i - \frac{n_{R} - \sigma_{R}}{2} \right) \frac{2\pi}{k}, 
    \eea
  For example the building block $\CB_{-}^{--}$ has $n_{L}=2$ and $n_{R}=-1$.
  This gives rise to a $(-\half)$ overall shift implying that the minimal puncture has $\sigma=-1$.
   
  For a theory associated to a cylinder there are two maximal punctures and a number of minimal punctures
  labeled, say by $p$.
  The above consideration leads to the conclusion that the signs and colors should satisfy the following relation
    \bea
    \sum_{p} \sigma_{p}^{(min)}
     =   (\sigma_{L} + n_{L} - 1) + (\sigma_{R} - n_{R} - 1),
         \label{U(1)tsphere}
    \eea
  up to $2k \mathfrak{n}$ ($\mathfrak{n} \in \IZ$), which is however absorbed into $n_{L}$ or $n_{R}$
  since this is defined mod $k$.
  The reason of the minus sign in front of $n_{R}$ is that the right puncture has negative orientation.
  The eq.~\eqref{U(1)tsphere} is interpreted as the conservation of the $U(1)_{t}$ curvature. 
  We can easily generalize this to the generic Riemann surface with maximal punctures labeled by $\alpha$
  and minimal puncture labeled by $p$ as
    \bea
    \sum_{p} \sigma_{p}^{(min)}
     =     \sum_{\alpha} (\sigma_{\alpha}^{(max)} + o_{\alpha} n_{\alpha} - 1),
           \label{U(1)t}
    \eea
  where $o_{\alpha}$ are the orientations of the maximal punctures.

\subsection{Linear and cyclic quivers}
\label{subsec:quiver}
  So far we saw a few examples of classes of chiral theories, orbifolded SQCD
  and (the orbifold of) a linear quiver as in Figure \ref{fig:branelinear}.
  In this subsection we construct more general $\CN=1$ theories by using the building blocks
  introduced above, still focusing on the building blocks where the signs of a puncture are the same.
  This will give a linear and cyclic quiver theories, and these are indeed the orbifolded versions of 
  the class of theories considered in \cite{Bah:2013aha}.

\paragraph{Gauging}
  Before studying generic quiver gauge theories, 
  we first consider the gauging of the flavor symmetry.
  The gauging already appeared in the beginning of Section \ref{sec:SQCD}. 
  However here we study it in a more generic setup.
  Let us suppose that we have a theory $\CT_{L}$ with a maximal puncture 
  with sign $\sigma$, negative orientation and color $n$,
  and another theory $\CT_{R}$ with a maximal puncture with sign $\tilde{\sigma} = - \sigma$, 
  positive orientation and color $n+2$.
  Since the anomaly $\tr \beta_{i+n-1} SU(N)_{i}^{2}$ of the first puncture cancels with that of the second puncture
  we have a ``good'' gauging of the diagonal $SU(N)^{k}$ symmetry,
  preserving the intrinsic symmetry.
  By this gauging we obtain a larger theory $\CT$.
  
  Let us see that the $U(1)_{t}$ condition in \eqref{U(1)t} is indeed consistent with the gauging.
  In order to have a correct behavior of the $U(1)_{t}$ curvature, the contribution from two punctures 
  we are gauging should be canceled.
  This is true for the above choice of $\sigma$'s, orientations and colors:
  $(\sigma + (-1)n - 1) + (-\sigma + (n+2) - 1) = 0$.
  Therefore by assuming the $U(1)_{t}$ conditions for the theories $\CT_{L}$ and $\CT_{R}$,
  we obtain the condition for $\CT$.
  
\begin{figure}[t]
	\centering 
	\includegraphics[width=10cm, bb=0 0 746 434]{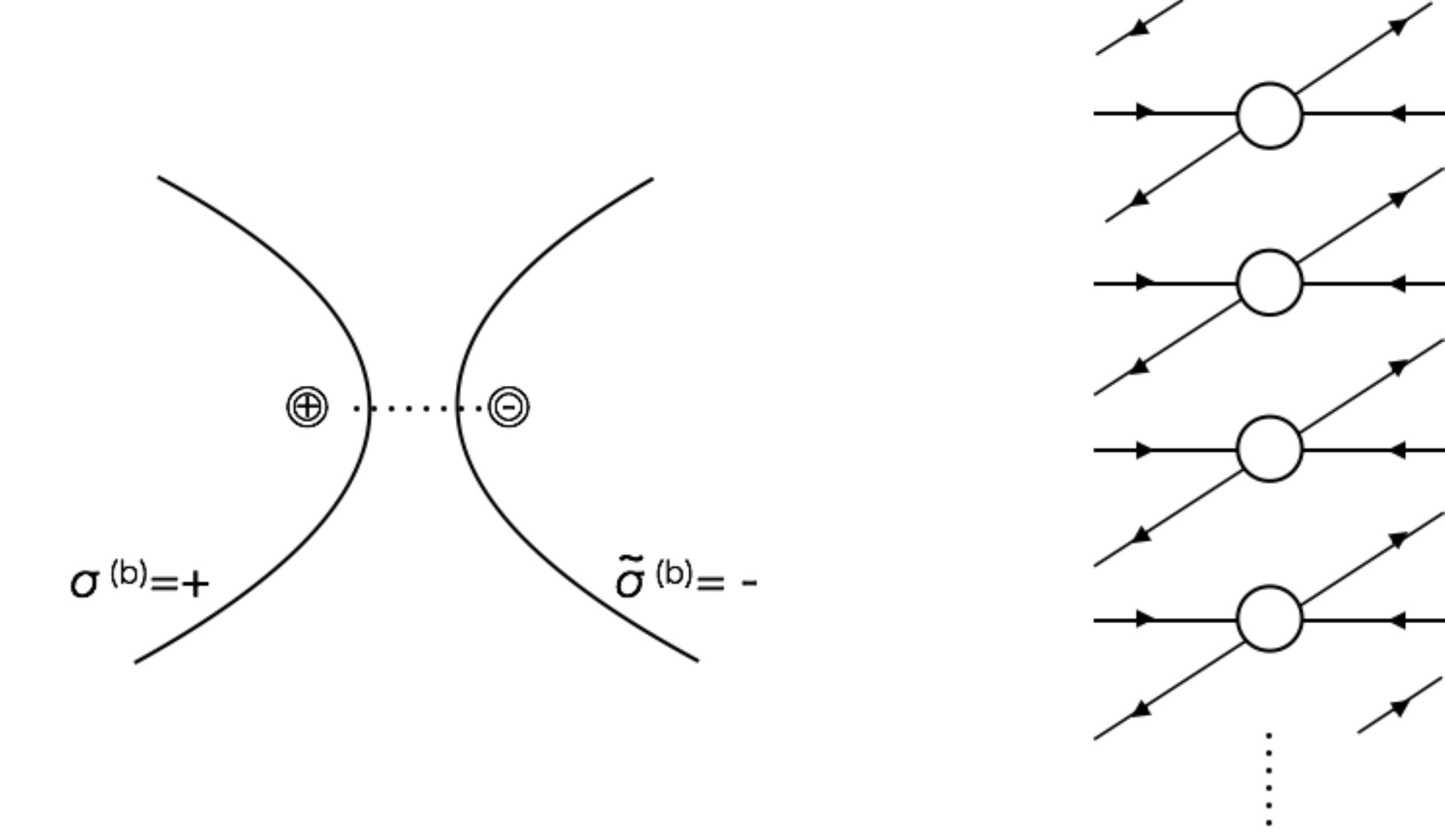}
	\caption{The gauging of the flavor symmetries associated to the gluing of the maximal punctures 
	         when $\sigma^{(b)} = +1$ and $\tilde{\sigma}^{(b)} = -1$.}
	\label{fig:gauging1}
    \end{figure}

  Let $\sigma^{(b)}$ and $\tilde{\sigma}^{(b)}$ be the signs of the pairs-of-pants
  to which the maximal punctures with signs $\sigma$ and $\tilde{\sigma}$ are attached respectively.
  Depending on these signs of pairs-of-pants, the gauging could be different.
  Since the overall sign difference corresponds to the overall flip of the arrows in the quiver, 
  we focus on the case with $\sigma= - \tilde{\sigma} = +1$.
  The first case is $\sigma^{(b)} = \sigma = +1$ and $\tilde{\sigma}^{(b)} = \tilde{\sigma}=-1$.
  This corresponds to the gauging which appears in the case of the orbifolded SQCD in Section \ref{sec:SQCD}.
  The gauging also adds the superpotential term which is bilinear in meson operators
    \bea
    W
     =    \tilde{M}_{i}^{~i+1} M_{i}^{~i+1},
    \eea
  where $M_{i}^{~i+1}$ and $\tilde{M}_{i}^{~i+1}$ are the mesons from two punctures.
  These meson operators are in turn bilinears of chiral multiplets, thus we obtain quartic couplings.
  See Figure \ref{fig:gauging1}.
  
\begin{figure}[t]
	\centering 
	\includegraphics[width=10cm, bb=0 0 746 434]{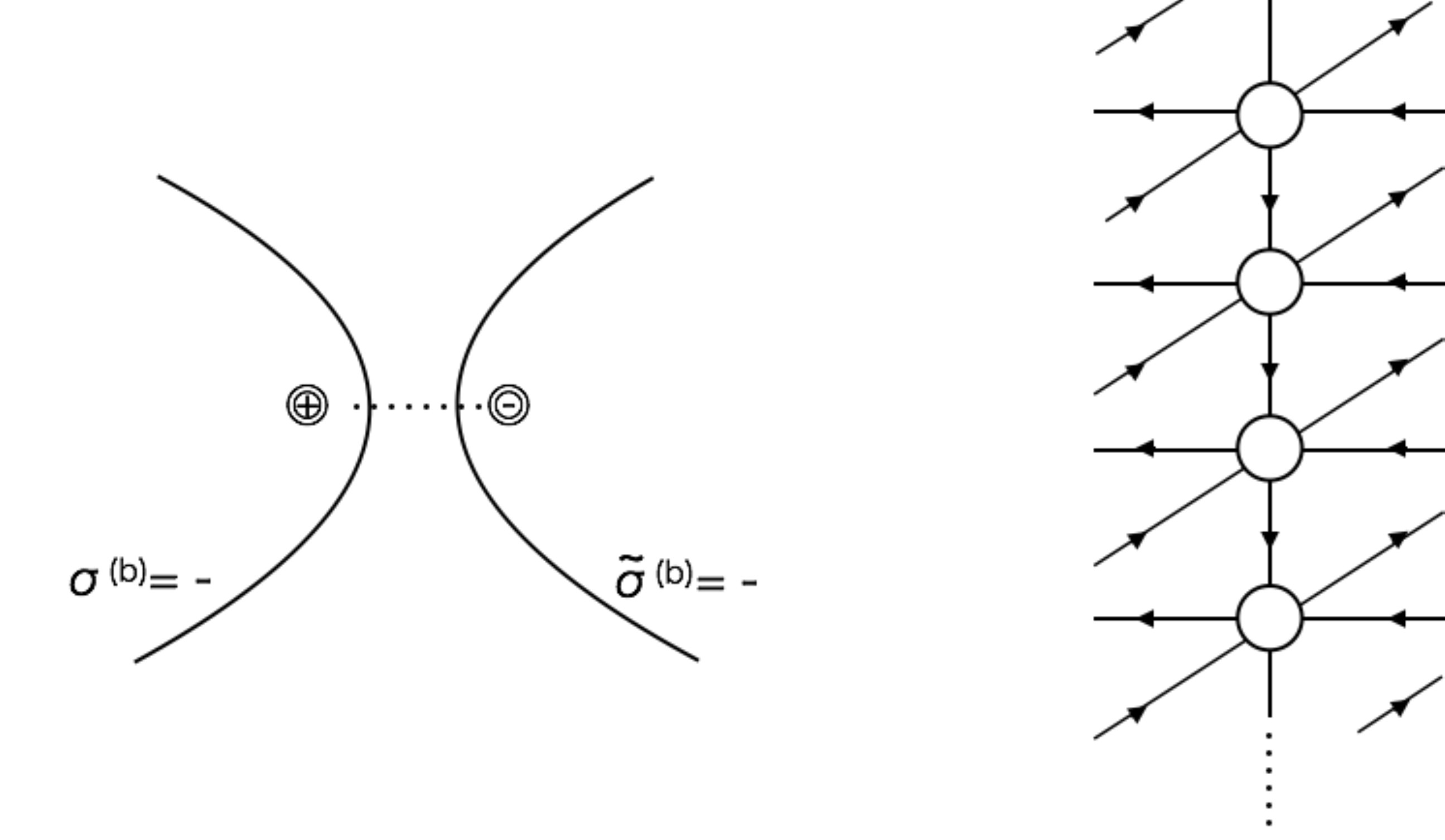}
	\caption{The gauging when $\sigma^{(b)} = \tilde{\sigma}^{(b)} = -1$.
	         In this case we additionally have bifundamental chiral multiplets.}
	\label{fig:gauging2}
    \end{figure}

  Let us then consider the case with $\sigma^{(b)}=\tilde{\sigma}^{(b)} = -1$.
  As shown in Figure \ref{fig:gauging2}, this corresponds to the introduction of 
  the bifundamental $\phi_{i}$ multiplets.
  The superpotential is again the bilinear form in the mesons, however in terms of $Q$ and $\phi$ it is cubic
    \bea
    W
     =     \tilde{M}_{i}^{~i+1} M_{i}^{~i+1}
     =     Q_{i+1} \tilde{Q}_{i+1} \phi_{i}.
    \eea
  This newly created cubic superpotential is for the triangles in the right hand side.
  The gauging in the case with $\sigma^{(b)}=\tilde{\sigma}^{(b)} = 1$ is similar, thus we skip it.
  Actually this gauging is the same as the one considered in \cite{Gaiotto:2015usa}.
  This can be understood from the six-dimensional perspective: 
  the same signs of the pairs-of-pants corresponds to considering only one line bundle,
  which gets back to the situation in \cite{Gaiotto:2015usa}, the orbifold of an $\CN=2$ theory.
  
\begin{figure}[t]
	\centering 
	\includegraphics[width=12cm, bb=0 0 970 436]{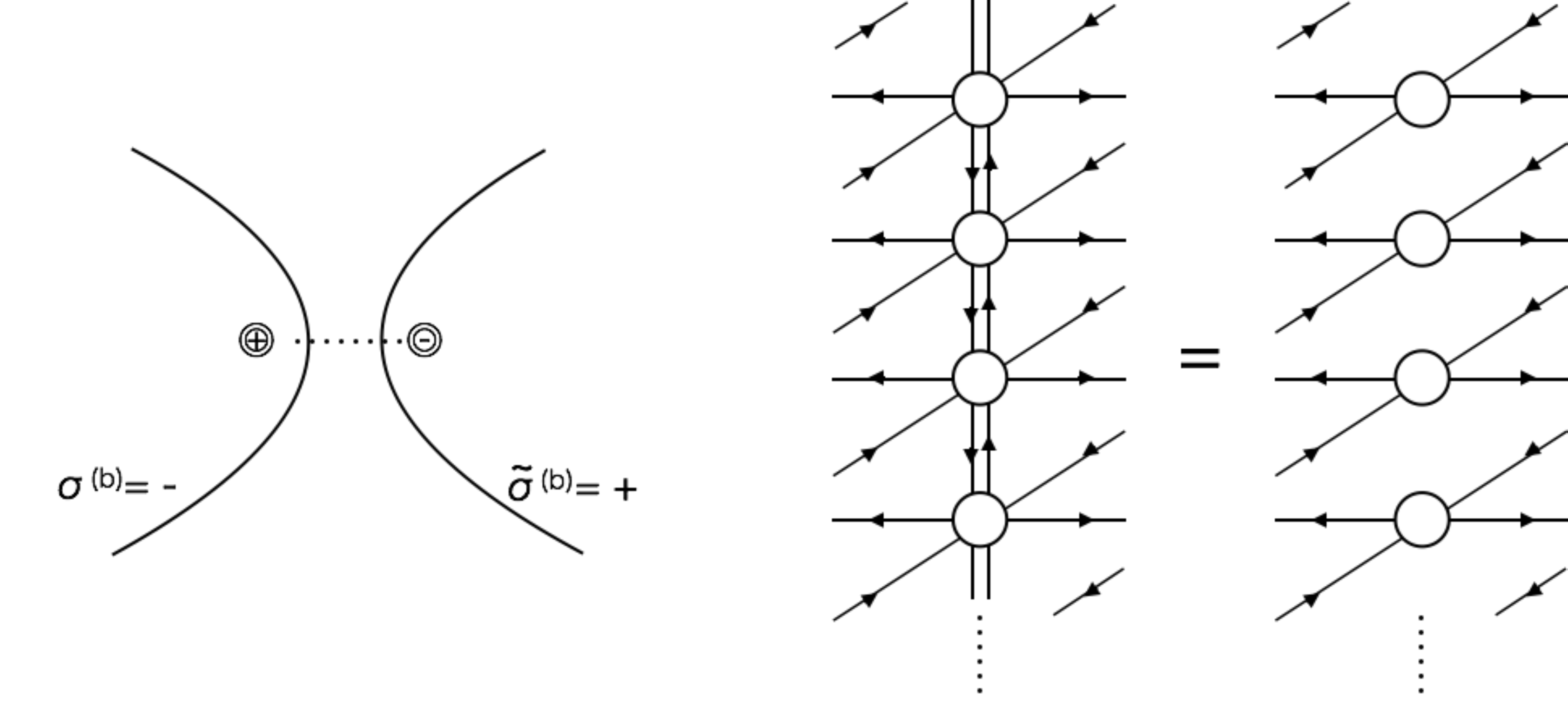}
	\caption{The gauging when $\sigma^{(b)} = -1$ and $\tilde{\sigma}^{(b)} = 1$.
	         This produces the mass terms for $\phi_{i}$ and $\tilde{\phi}_{i}$.}
	\label{fig:gauging3}
    \end{figure}

  Finally the gauging in the case with $\sigma^{(b)} = - \tilde{\sigma}^{(b)} = -1$.
  As depicted in Figure \ref{fig:gauging3}, there are two sets of bifundamental multiplets, 
  $\phi$ and $\tilde{\phi}_{i}$.
  The superpotential is 
    \bea
    W
     =     \phi_{i} \tilde{\phi}_{i} +  \tilde{Q}_{i+1} Q_{i+1} \phi_{i} +  q_{i+1} \tilde{q}_{i} \tilde{\phi}_{i},
    \eea
  where the first term is from $\tilde{M}_{i}^{~i+1} M_{i}^{~i+1}$, and the second and the third terms from 
  the building blocks on the left and the right.
  Since $\phi_{i}$ and $\tilde{\phi}_{i}$ are massive, after integrating them out, 
  we are left with the quartic superpotential, depicted in the right of Figure \ref{fig:gauging3}.
  There are no bifundamental multiplets and the result is the same as the first gauging.
    
  In summary, what we just observed here is that the gauging (or gluing of pairs-of-pants) is puncture-independent:
  when the signs of the pairs-of-pants are the same, we just gauge the flavor symmetry;
  when the signs of the pairs-of-pants are different, we gauge and add the bifudamental multiplets.
  Here we use the building blocks which are studied in subsection \ref{subsec:bb}.
  However the analysis is completely generic when the building blocks are of the other type.
  Note that the gauging from the point of view of the superconformal index has been argued in \cite{Gaiotto:2015usa}.
  
%
\paragraph{Linear quiver}
  Let us now consider the Riemann sphere with two maximal punctures 
  with ($n_{L}, \sigma_{L}$) and ($n_{R},\sigma_{R}$)
  and $n-2$ minimal punctures with $\sigma^{(min)}_{p}$ ($p=1,\ldots,n-2$).
  We define $\sigma_{tot}=\sum_{p} \sigma_{p}^{(min)}$,
  and denote the class of theories associated to the Riemann sphere 
  as $\CL_{n-2,\sigma_{tot}}^{(n_{L}, \sigma_{L}),(n_{R}, \sigma_{R})}$.
  The sphere is decomposed into $n-2$ pairs-of-pants, 
  and each pants decomposition corresponds to a UV gauge theory description.
  We propose that all the theories from all possible pants decompositions flow to the same IR fixed point.
  We will restrict to pairs-of-pants which are studied above, corresponding to the subclass 
  where the signs of the minimal punctures $\sigma_{p}^{(min)}$ are equal to $\sigma_{p}^{(b)}$, 
  the signs of the pairs-of-pants.
  
\begin{figure}[t]
	\centering 
	\includegraphics[width=8cm, bb=0 0 622 566]{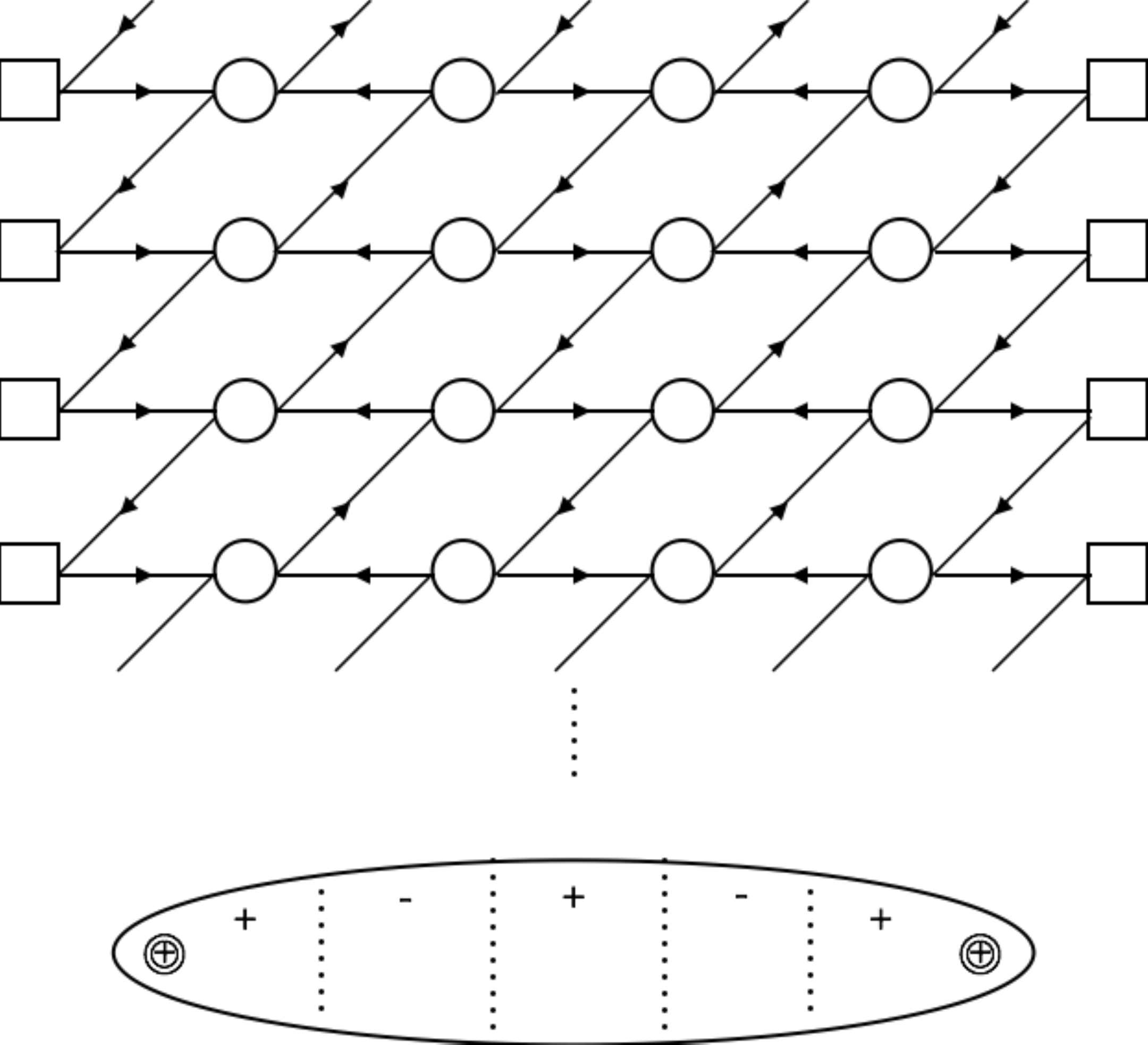}
	\caption{A linear quiver theory in the class $\CL_{5,1}^{(n+1,+),(n,+)}$.}
	\label{fig:lquiver1}
    \end{figure}

  The rule of drawing the quiver diagram follows the usual brane tiling prescription \cite{Franco:2005rj}.
  The quiver is cyclic in the vertical direction and has a linear shape in the horizontal direction. 
  The gauging has been already studied.
  At each end of the quiver, if the sign of the maximal puncture is different from the sign of its pair-of-pants, 
  we may add vertical bifundamental chiral multiplets.
  An example corresponding to $n=7$ and $\sigma=\tilde{\sigma}=+$,
  $\sigma^{1} = - \sigma^{2} = \sigma^{3} = - \sigma^{4} = \sigma^{5} = +$ is depicted in Figure \ref{fig:lquiver1}.
  This is in a class $\CL_{5,1}^{(n+1,+),(n,+)}$.
  Since the signs of the adjacent minimal punctures are opposite, we get gaugings without bifundamentals. 
  
\begin{figure}[t]
	\centering 
	\includegraphics[width=8cm, bb=0 0 622 566]{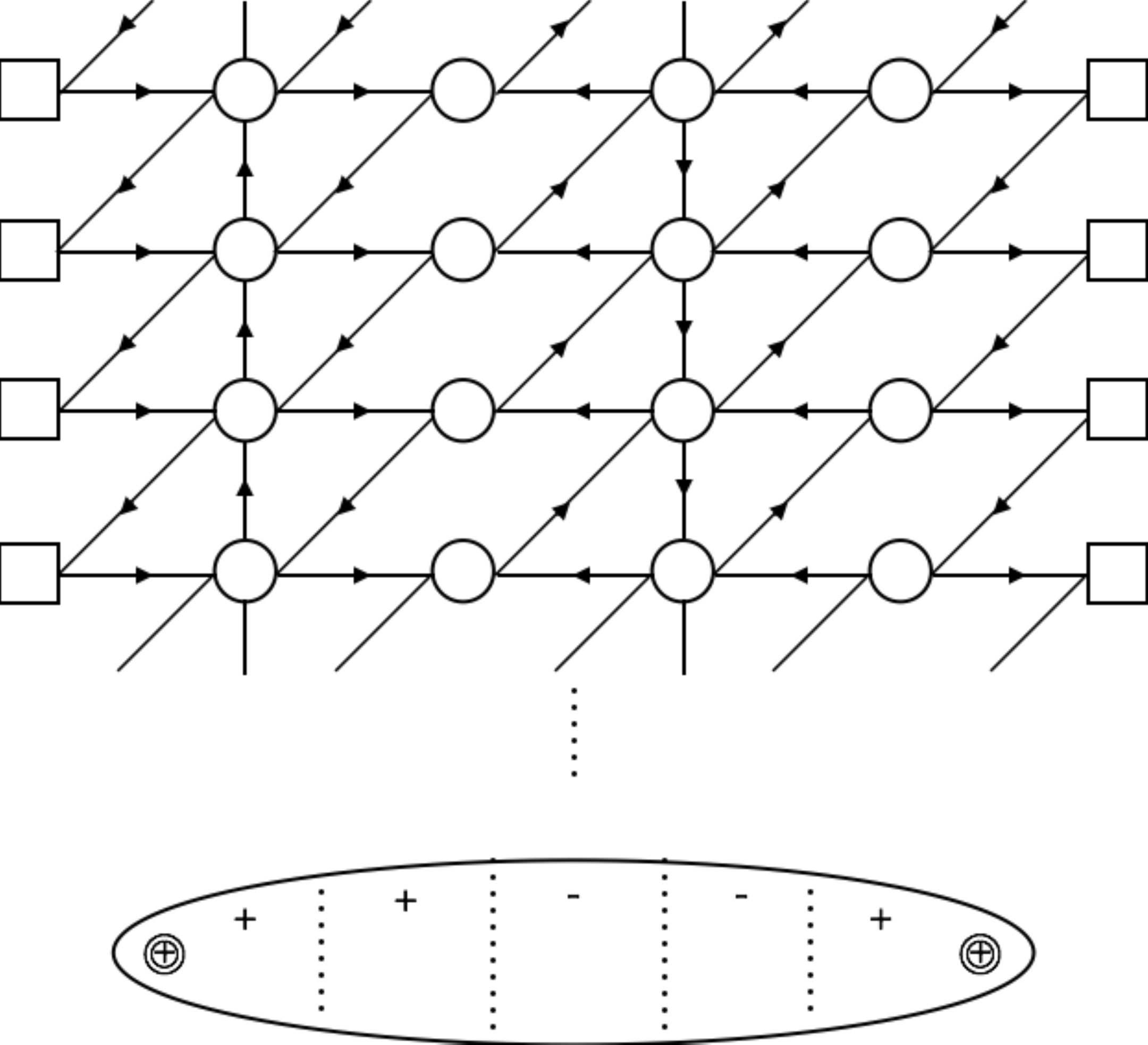}
	\caption{Another quiver theory in the class $\CL_{5,1}^{(n+1,+),(n,+)}$.
	         This is obtained by Seiberg dualities to the second set of the vertical nodes
	         in the quiver in Figure \ref{fig:lquiver1}.}
	\label{fig:lquiver2}
    \end{figure}

  One can think of the action of the Seiberg duality on a vertical set of gauge nodes as an 
  exchange of the minimal punctures and the corresponding pairs-of-pants.
  For example let us Seiberg-dualize the second (from the left) vertical set of gauge nodes 
  in Figure \ref{fig:lquiver1}.
  After the duality we get sets of vertical bifundamentals for the first and the third gauge nodes 
  (namely the adjacent gauge nodes to the dualized ones) as depicted in Figure \ref{fig:lquiver2}, 
  and this is indeed the theory associated to the order 
  of the minimal punctures $\{ \sigma_{1}, \sigma_{3}, \sigma_{2}, \sigma_{4}, \sigma_{5} \}$.
    
  Finally the difference of the colors of the two maximal punctures are written in terms of 
  the signs by using the condition for the $U(1)_{t}$ curvature \eqref{U(1)t}.

\paragraph{Cyclic quiver}
  One can easily construct a quiver gauge theory associated to a torus with $n$ minimal punctures
  with $\sigma_{p}$ ($p=1,2,\ldots,n$).
  Let us denote this class of theories as $\CC_{n,\sigma_{tot}}$ where $\sigma_{tot} = \sum_{p} \sigma_{p}$.
  This theory is periodic both in the horizontal and vertical directions.
  A condition for $\sigma_{tot}$ comes from the $U(1)_{t}$ curvature:
    \bea
    \sigma_{tot}
     =     2k\mathfrak{n} ,
           \label{cond}
    \eea
  where $\mathfrak{n} \in \IZ$.
  This condition ensures the existence of all the intrinsic $U(1)$ symmetries.
  
\begin{figure}[t]
	\centering 
	\includegraphics[width=7cm, bb=0 0 560 422]{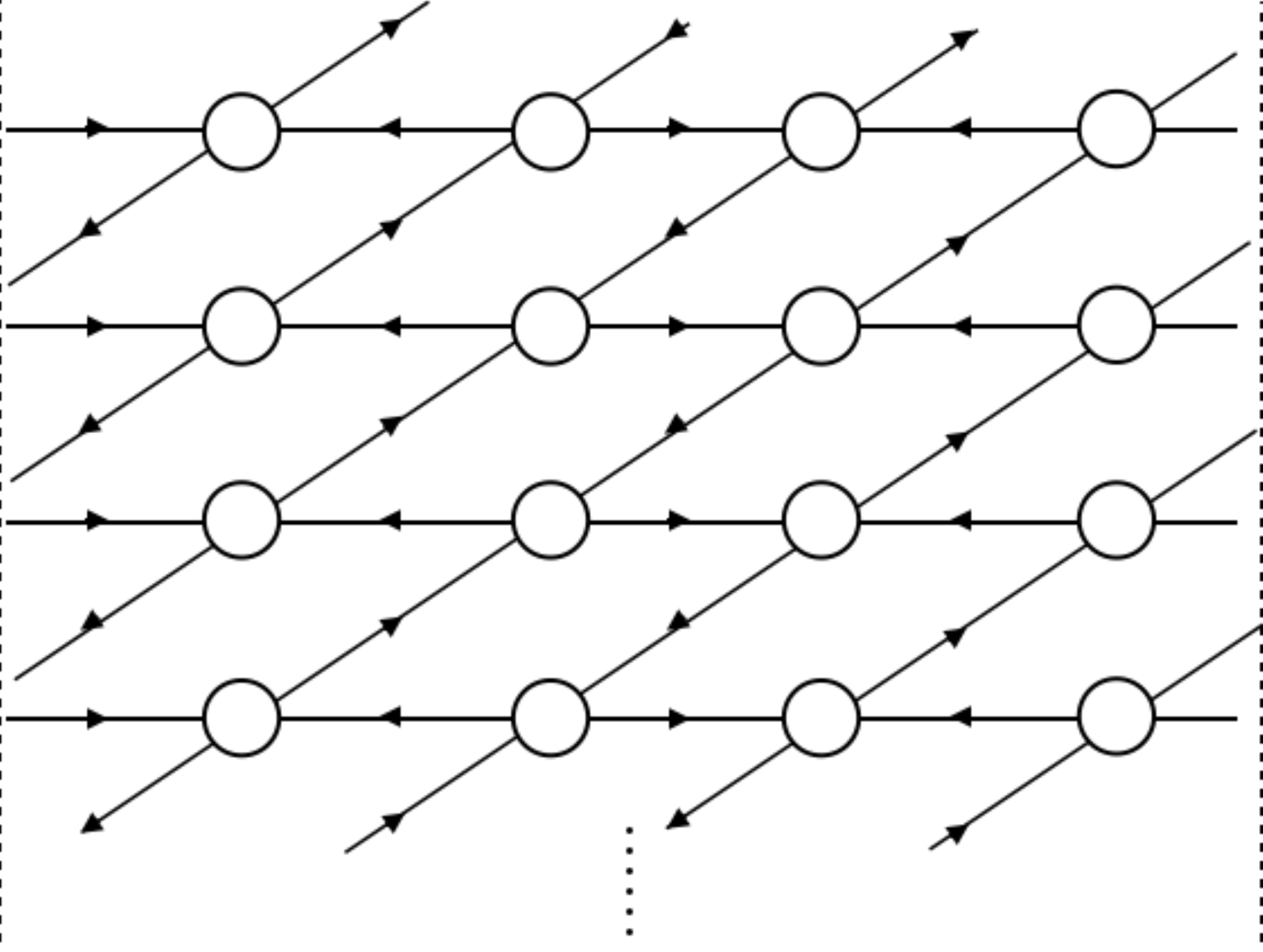}
	\caption{A cyclic quiver in the class $\CC_{4,0}$. Cyclic both in the horizontal and vertical directions.}
	\label{fig:cquiver}
    \end{figure}

  The cyclic quiver drawn in Figure \ref{fig:cquiver} is an example of the theory 
  in the class $\CC_{4,0}$. 
  This is the same quiver as the orbifold of the conifold theory in \cite{Uranga:1998vf}
  with an orbifold action of $\mathbb{Z}_{2} \times \mathbb{Z}_{k}$.

\begin{figure}[t]
	\centering 
	\includegraphics[width=11cm, bb=0 0 902 463]{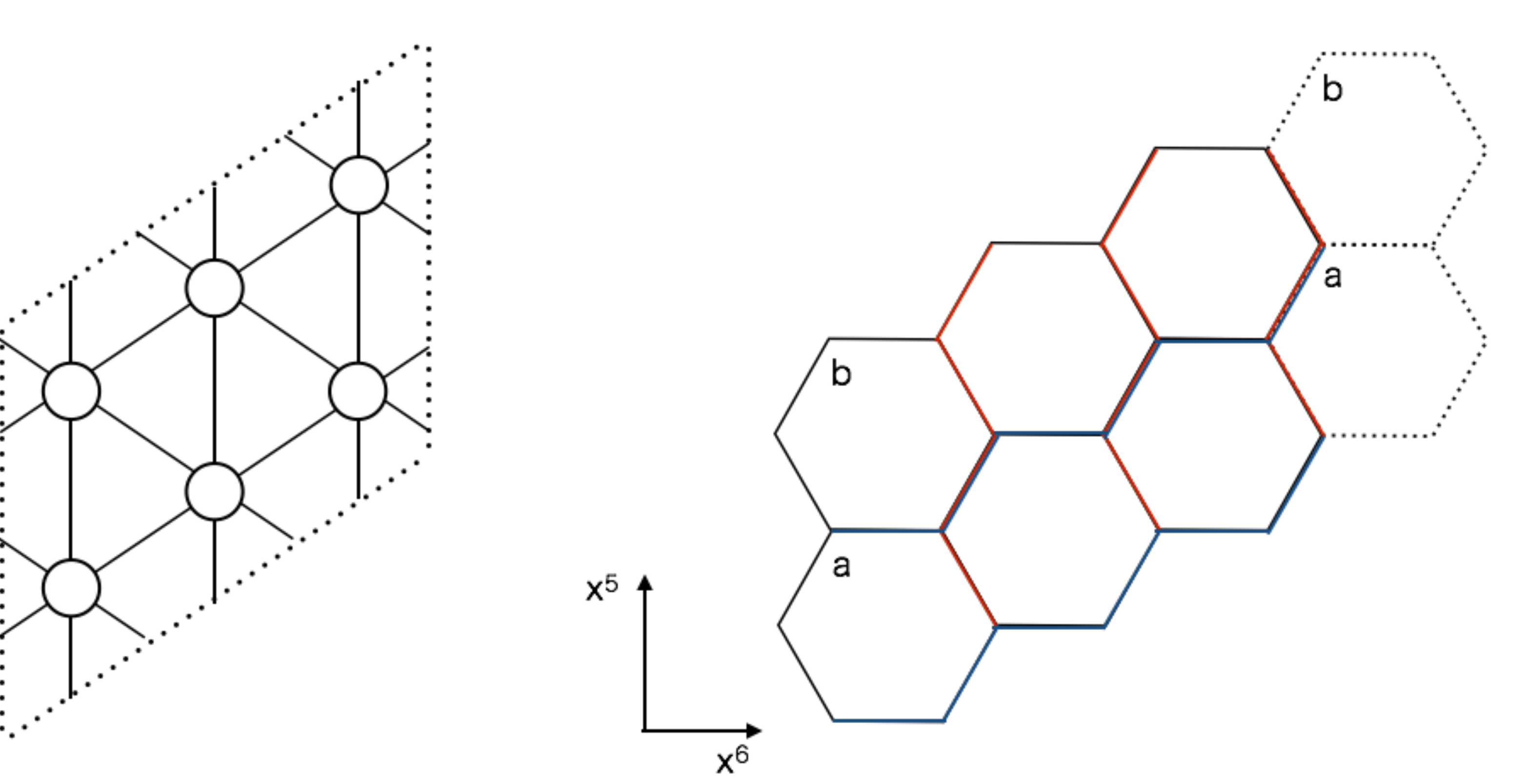}
	\caption{Left: a quiver gauge theory in the class $\CC_{3,3}$.
	         Right: a corresponding brane tiling.
	         These are periodic in the vertical 
	         and the horizontal directions. (The tiles with $a$ ($b$) are identified.)
	         Zig-zag paths in the red lines are associated to minimal punctures,
	         and thus to NS5-branes.
	         Zig-zag paths in the blue lines corresponds to NS5$''$-branes.}
	\label{fig:tiling}
    \end{figure}

  The cyclic quiver theories are included in brane tiling family \cite{Franco:2005rj}.
  The brane tiling consisting of D5-, NS5, and NS5$''$-branes as in Section  \ref{subsec:brane} is drawn on 
  $T^{2}$ in the $x^{5}$ and $x^{6}$ directions in this case.
  Note that this $T^{2}$ is different from the torus (spanned by $x^{6}$ and M-circle $x^{10}$) 
  on which we compactify the six-dimensional theory.
  Let us consider, {\it e.g.}, a quiver theory in the class $\CC_{3,3}$ obtained from the $\IZ_{2}$ orbifold 
  as in Figure \ref{fig:tiling}.
  (This does not satisfies \eqref{cond}, thus some of the intrinsic symmetries are broken.)
  The brane tiling is depicted in the right in Figure \ref{fig:tiling}.
  
  A remarkable feature is the symmetry under the exchange of the roles of the $x^{5}$ and $x^{6}$ directions 
  because in Type IIB there is no distinction between NS5- and NS5$''$-branes.
  However this causes a drastic change of the view of the six-dimensional theory:
  one can see this theory as a compactification of the six-dimensional theory 
  associated to the $\IC^{2}/\IZ_{3}$ orbifold on a torus with two punctures 
  where the torus is in the $x^{5}$ and $x^{10}$ directions.
  From the global symmetry point of view, this is the exchange of the zig-zag paths.
  The three paths denoted by the red lines correspond to three minimal punctures,
  and two paths by the blue lines correspond to the orbifold in the original picture.
  The roles are exchanged by going to the other picture.
  In general we can state that the same quiver theory can be obtained from
  the six-dimensional theory associated to $\IC^{2}/\IZ_{k}$ 
  on a torus with $n$ minimal punctures with the same signs,
  and from the one associated to $\IC^{2}/\IZ_{n}$ on a torus with $k$ minimal punctures.
  
  The brane tiling includes various quiver theories which cannot be constructed from the building blocks
  in the way we explained in this section.
  However, the above observation may open up a direction to reach such generic quivers
  by generalizing the way to treat minimal punctures.
  The symmetric structure in $x^{5}$ and $x^{6}$ directions, or in other words 
  the zig-zag paths which correspond to $U(1)$ symmetries, suggests to define
  an NS5-brane (corresponding to a zig-zag path wrapping $(1,0)$ cycle in the $x^{5}$ direction) 
  as $(1,0)$ minimal puncture,
  and to define an NS5$''$-brane (corresponding to a zig-zag path wrapping $(0,1)$ cycle in the $x^{6}$ direction) 
  as $(0,1)$ minimal puncture.
  The $(1,0)$ puncture is exactly the one which we called as a minimal puncture before.
  It is then natural to have a mixed type of puncture, namely $(x,y)$ with $x\neq0$ and $y\neq0$,
  such that the corresponding zig-zag path wraps $(x,y)$ cycle on the torus.
  This leads to more general quiver theories.
  It is obscure how to define this type of punctures from the six-dimensional point of view. 
  We leave this problem as a future work.

\section{Higgsing}
\label{sec:higgs}
  In this section we consider the Higgsing of the chiral theories by giving vevs to the baryon operators.
  This corresponds to completely closing the minimal punctures, 
  with the introduction of the discrete curvatures of $U(1)_{\beta_{i}}$ or $U(1)_{\gamma_{i}}$ and $U(1)_{t}$.
  
\begin{figure}[t]
	\centering 
	\includegraphics[width=11cm, bb=0 0 802 582]{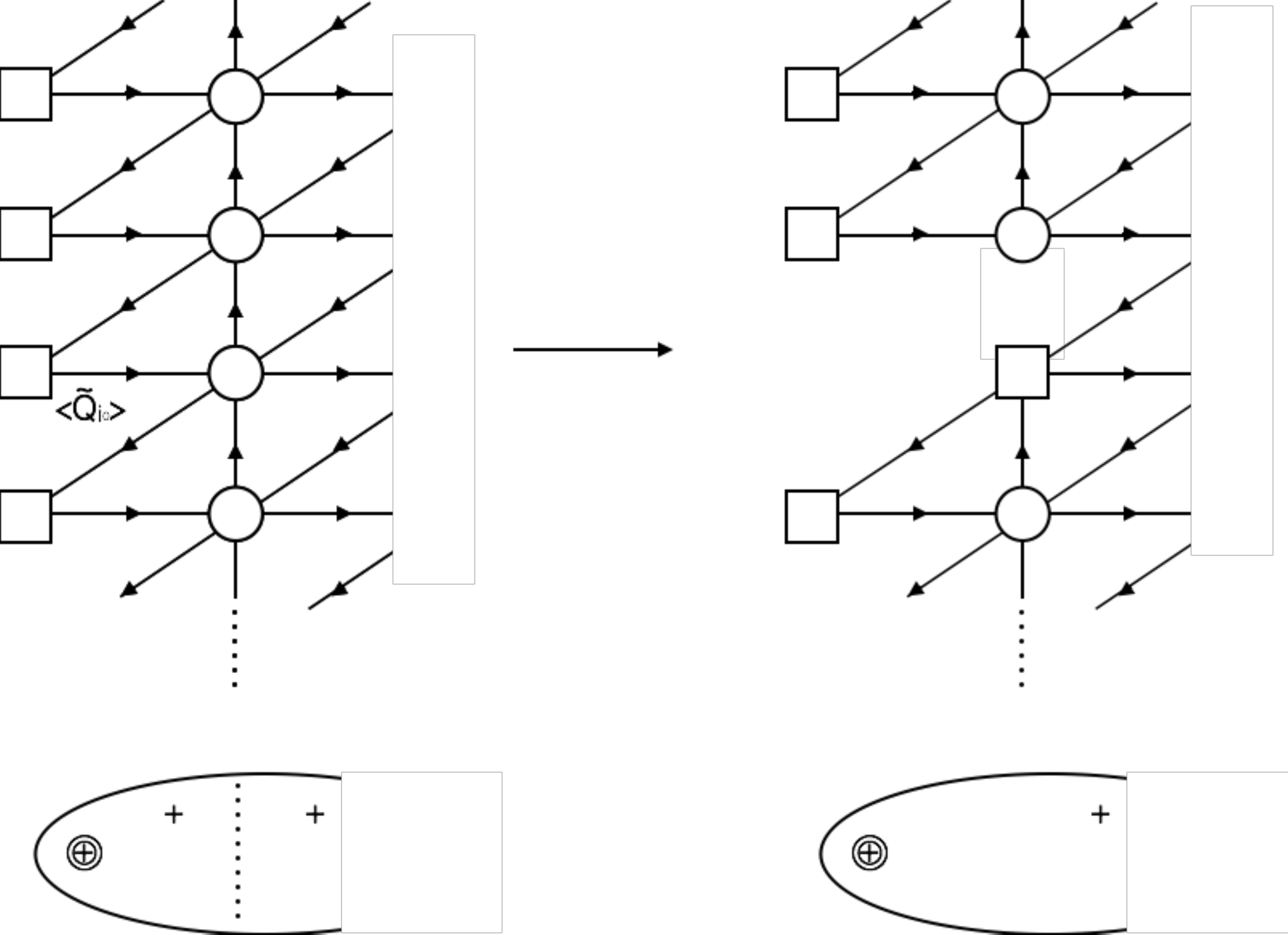}
	\caption{Higgsing by a vev to a baryon $\tilde{Q}_{i_{0}}^{N}$.
	         This completely closes the minimal puncture.}
	\label{fig:Higgs1}
    \end{figure}

  Let us first recall the Higgsing of the theories associated to the Riemann surface 
  with punctures where all the signs are the same \cite{Gaiotto:2015usa}.
  We consider the duality frame where the theory is described as in Figure \ref{fig:Higgs1}.
  There are $2k$ different ways to give the vevs to $2k$ different baryons, $Q_{i}^{N}$ and $\tilde{Q}_{i}^{N}$
  with charges $R^{\frac{N}{2}}t^{\frac{N}{2}}\beta_{i}^{N}\alpha^{N}$ 
  and $R^{\frac{N}{2}}t^{\frac{N}{2}}\gamma_{i}^{-N}\alpha^{-N}$.
  These correspond to the different ways to close the left-most minimal puncture.
    
  Let us consider the vev to $\tilde{Q}_{i=i_{0}}^{N}$.
  More precisely we give equal vevs to the diagonal components $(\tilde{Q}_{i_{0}})_{\alpha}^{~\alpha}$
  with $R^{\half} t^{\half} \gamma_{i_{0}}^{-1} \alpha^{-1}$.
  Due to the vevs the $SU(N)_{i_{0}}$ gauge group is Higgsed, and get an $SU(N)$ flavor symmetry
  which is the diagonal part of the gauge $SU(N)_{i_{0}}$ and flavor $SU(N)_{i_{0}}$.
  The cubic coupling makes $Q_{i_{0}}$ and $\phi_{i_{0}}$ massive.
  Thus we get a quiver as in Figure \ref{fig:Higgs1}.
  
  The charges of the chiral multiplets under the global symmetries are shifted by the vevs.
  This is determined by setting the vevs to be neutral under all the symmetry.
  In this case we set $\alpha=R^{\half} t^{\half} \gamma_{i_{0}}^{-1}$.
  In other words, we shift the $U(1)$ charges as: $U(1)_{R} \rightarrow U(1)_{R} + \half U(1)_{\alpha}$,
  $U(1)_{t} \rightarrow U(1)_{t} + \half U(1)_{\alpha}$, 
  and $U(1)_{\gamma_{i_{0}}} \rightarrow U(1)_{\gamma_{i_{0}}}- U(1)_{\alpha}$.
  After the shift of the charges, we have
    \bea
    Q_{i}: R t \beta_{i} \gamma_{i_{0}}^{-1}, ~~~
    \tilde{Q}_{i}: \gamma_{i_{0}} \gamma_{i}^{-1}~(i\neq i_{0}), ~~~
    \phi_{i}: R t^{-1} \beta_{i}^{-1} \gamma_{i}.
    \eea
  One can check that the $\tr \beta_{i} SU(N)_{i}^{2}$ (and $\tr \gamma_{i} SU(N)_{i}^{2}$) anomalies 
  are not changed. 
  Thus the sign and the color of the maximal puncture is kept intact, 
  while the minimal puncture with $U(1)_{\alpha}$ has gone.
  This theory is associated to the same Riemann surface but with one less minimal punctures
  compared to the theory before the Higssing.
  
  For instance let us consider the class of theories $\CL^{++}_{n-2,n-2}$ considered in section \ref{subsec:quiver}.
  On a dual frame, we have a description by a linear quiver.
  It is obvious that the theory obtained by Higgsing as above is different from the linear quiver 
  in the class $\CL^{++}_{n-3,n-3}$ associated to the same sphere with one less minimal puncture.
  This difference is due to the introduction of
  minus one unit of $U(1)_{\gamma_{i_{0}}}$ discrete curvature \cite{Gaiotto:2015usa} 
  and one unit of $U(1)_{t}$ discrete curvature.
  This is because we gave the vev to the field with $t^{\half} \gamma_{i_{0}}^{-1}$.
  Indeed, the $U(1)_{t}$ curvature can be computed by using \eqref{U(1)t} as follows.
  Before the Higgsing we had the condition $n-2 = n_{L}-n_{R}$ where $n_{L}$ ($n_{R}$) is the color of the 
  left (right) maximal puncture.
  The Higgsing does not change the colors, but reduces the number of the minimal punctures by one.
  Therefore the Riemann surface should have one unit of the $U(1)_{t}$ curvature after the Higgsing.
  We see this in the next section by using the anomaly coefficients.
  
  We could give vevs to the baryon with fugacities $\beta^{N}_{i}$.
  This gives a different theory with one unit of discrete curvatures of $U(1)_{\beta_{i}}$ and $U(1)_{t}$.

\begin{figure}[t]
	\centering 
	\includegraphics[width=14cm, bb=0 0 1348 585]{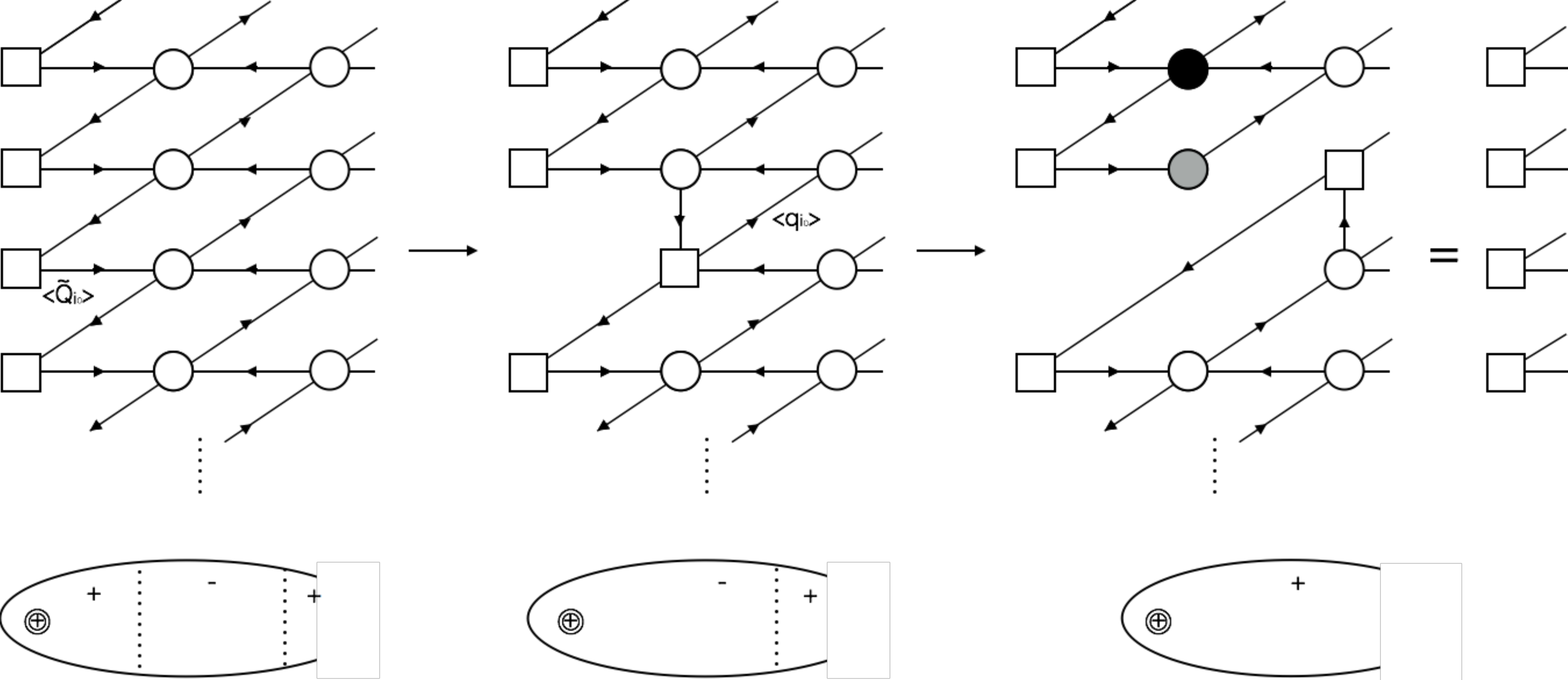}
	\caption{Left to middle: the Higgsing by the vev to the baryon $\tilde{Q}_{i_{0}}^{N}$.
	         This completely closes the minimal puncture with $\sigma=+$.
	         Middle to right: the Higgsing by the vev to the baryon $q_{i_{0}}^{N}$
	         which closes another minimal puncture with $\sigma=-1$.
	         The shaded node now has $N_{f}=N$, and is confined.
	         This induces the cascade of the confinement.}
	\label{fig:Higgs2}
    \end{figure}

  Let us then consider the Higgsing in the theories associated to a Riemann surface with punctures 
  with different signs.
  Suppose a dual frame where the quiver is as in Figure \ref{fig:Higgs2},
  and give a vev to a baryon $\tilde{Q}_{i_{0}}^{N}$ as above.
  In this case no multiplet becomes massive and the Higgsing gives rise to the quiver 
  in the middle in Figure \ref{fig:Higgs2}.
  We interpret this as the theory associated to a Riemann surface with one less minimal puncture with $\sigma=+$
  but with minus one unit of the discrete curvature of $U(1)_{\gamma_{i_{0}}}$ and one unit of $U(1)_{t}$.
  Again there is no change in the $\tr \beta_{i} SU(N)_{i}^{2}$ and $\tr \gamma_{i} SU(N)_{i}^{2}$ anomalies.
  Thus the introduction of the discrete $U(1)_{t}$ curvature can be understood from \eqref{U(1)t}.
  
  One can continue the Higging to give a vev to a baryon in the next building block.
  This will add further discrete curvatures of $U(1)_{\gamma}$ or $U(1)_{\beta}$ and $U(1)_{t}$.
  Note that in this case the $U(1)_{\gamma}$ (or $U(1)_{\beta}$) charges of the bifundamentals
  in the next building block are of opposite signs to those of the first building block. 
  An interesting observation is that when we give a vev to $q_{i_{0}}$ with charge $t^{-\half}\gamma_{i_{0}}$ 
  as in the middle in Figure \ref{fig:Higgs2},
  this adds plus one unit of the $U(1)_{\gamma_{i_{0}}}$ curvature and minus one unit of $U(1)_{t}$.
  Therefore we should get back to the theory associated to
  a Riemann surface with two less minimal punctures with $\sigma=+$ and $\sigma=-$ without any curvature.
  Let us see this is indeed the case here.
  After the Higgsing from the middle to the right in Figure \ref{fig:Higgs2}, 
  we get a gauge node with $N_{f}=N$ flavors (the grey colored node).
  Thus at those nodes the theory is confined in the infrared \cite{Seiberg:1994bz}, 
  and described by the gauge invariant operators, baryons and mesons, with the constraint
  $\det M - B \tilde{B} = \Lambda^{2N}$.
  Let us suppose that $\langle M \rangle = \Lambda^{2}$ and $\langle B \rangle = \langle \tilde{B} \rangle = 0$.
  This induces the Higgsing further, and only the diagonal part of the two $SU(N)$ factors
  which are attached to the confined $SU(N)$ survive.
  At the same time this gives masses to two bifundamental multiplets of the (black colored) gauge node 
  and the Higgsed nodes.
  Thus the number of flavors at the black gauge node becomes $N_{f}=N$.
  This induces the another confinement.
  Like this all the remaining gauge nodes confine, and we finally obtain the $SU(N)^{k}$ flavor symmetry.
  
  We should note that the vev's of all baryons other than $\tilde{Q}_{i_{0}}^{N}$ and $q_{i_{0}}^{N}$ are assumed to be 0.
  As argued in \cite{Gaiotto:2015usa}, this is achieved by adding multiplets which quadratically couples to 
  these baryons.
We will use the same terms here to get the final result.  
  
\begin{figure}[t]
	\centering 
	\includegraphics[width=15cm, bb=0 0 1449 624]{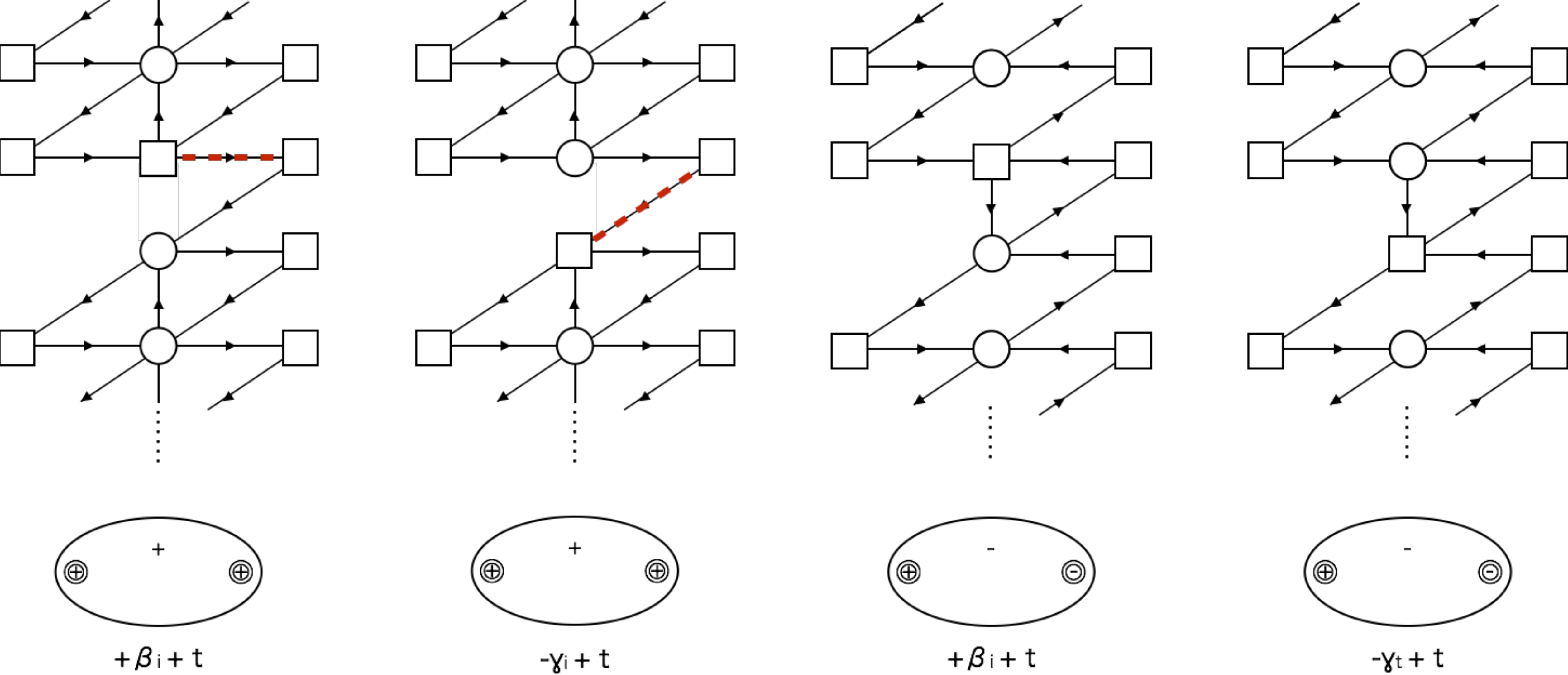}
	\caption{The building blocks (pairs-of-pants) 
	         in presence of the discrete curvatures of $U(1)_{\beta}$ or $U(1)_{\gamma}$ and $U(1)_{t}$.}
	\label{fig:buildingblock6}
    \end{figure}

  The Higgsings by giving vevs to different baryons can be analyzed in a similar way.
  This produces $2k$ inequivalent building blocks with discrete curvatures of 
  $U(1)_{\beta}$ or $U(1)_{\gamma}$ and $U(1)_{t}$.
  Altogether, we find building blocks as summarized in Figure \ref{fig:buildingblock6},
  where the flavor node which has changed the position is always $SU(N)_{i_{0}}$. 
  For the left two building blocks, we have two maximal punctures with $o=+1$, $n=1$ and $o=-1$, $n=-1$.
  All the signs are the same as $\sigma=+$.
  For the right two building blocks, we have two maximal punctures with $o=+1$, $n=1$ and $o=-1$, $n=-1$.

\section{Anomaly coefficients and central charges}
\label{sec:anomaly}
  In this section we consider the 't Hooft anomaly coefficients and central charges of the class of chiral theories
  studied in the previous sections.

\subsection{Anomalies of chiral theories of class $\CS$}
  We consider the 't Hooft anomalies of $U(1)_{R}$ and $U(1)_{t}$ symmetries of the class of chiral theories,
  and see a geometric interpretation of them.
  Since there is a similarity between the $\CN=1$ class $\CS$ theories 
  and the class of chiral theories studied in this paper,
  we first review shortly the form of the anomaly coefficients of the former theories.
  
  Let us focus on two $U(1)_{R}$ and $U(1)_{\CF}$ symmetries.
  It is known that the following combinations of these $U(1)$'s are convenient to study:
    \bea
    J_{\pm}
     =     R \pm \CF,
    \eea
  where $R$, $\CF$ and $J_{\pm}$ are the generators of $U(1)_{R}$, $U(1)_{\CF}$ and $U(1)_{J_{\pm}}$.
  For this class of theories the relation $\tr J_{\pm} = \tr J_{\pm}^{3}$ is always satisfied.
  Then the anomaly coefficients are the following:
    \bea
    \Tr J_{+}
     =     \Tr J_{+}^{3} 
    &=&    q(N-1) + 2 \sum_{i (\sigma = -1)} (n_{v}(Y_{i}) - n_{h}(Y_{i})), 
           \label{J+} \\
    \Tr J_{-}
     =     \Tr J_{-}^{3} 
    &=&    p(N-1) + 2 \sum_{i (\sigma = +1)} (n_{v}(Y_{i}) - n_{h}(Y_{i})),
           \label{J-} \\
    \Tr J_{+}^{2} J_{-}
    &=&    \frac{p}{3} (4N^{3} - N - 3) + 2 \sum_{i (\sigma = +1)} n_{v}(Y_{i}),
           \nonumber \\
    \Tr J_{+} J_{-}^{2}
    &=&    \frac{q}{3} (4N^{3} - N - 3) + 2 \sum_{i (\sigma = -1)} n_{v}(Y_{i}),
           \label{anomaliesN=1}
    \eea
  where $i$ labels all the punctures, and $p$ and $q$ are the degrees of the line bundles.
  The first terms on the rhs correspond to the contributions from the Riemann surface
  and the second terms correspond to the contributions from the punctures.
  The puncture is specified by a Young diagram and its contribution is given by 
    \bea
    n_{v}(Y)
    &=&    \sum_{k=2}^{N} (2k-1) p_{k} - \frac{1}{6} (4N^{3} - N - 3),
           \nonumber \\
    n_{v}(Y) - n_{h}(Y)
    &=&  - \frac{1}{2} (\sum_{r} \ell_{r}^{2} - 1 ),
           \label{N=1puncture}
    \eea
  where $p_{k}$ is associated to $k$-th box and given by $p_{k} = k-h$ 
  where $h$ is the height of the box.
  For the maximal puncture $p_{k} = (0,1,2,\ldots,N-1)$ thus we get
  $n_h (Y_{\textrm{max}}) = 0$ and $n_v (Y_{\textrm{max}}) = - \frac{1}{2} (N^{2}-1)$.
  The minimal puncture gives 
  $n_h (Y_{\textrm{min}})= - \frac{1}{6}(4N^{3} - 6N^{2} - 4N)$ and 
  $n_v (Y_{\textrm{min}})= - \frac{1}{6} (4N^{3} - 6N^{2} - N + 3)$. 
  
  Now let us study the class of chiral theories.
  In analogy with the above discussion we introduced the two linear combinations of the $U(1)$ symmetries:
    \bea
    J_{\pm}
     =     R \pm T,
    \eea
  where $R$ and $T$ are the generators of the $U(1)_{R}$ and $U(1)_{t}$ symmetries.
  One can see that for the building block $\CB^{++}_{+}$, the bifundamental chiral multiplets have
  $J_{\pm}$ charges $(J_{+},J_{-}) = (1,0)$.
  For the building block $\CB^{-+}_{+}$, 
  the bifundamentals $Q$ and $\tilde{Q}$ have charges $(J_{+},J_{-}) = (1,0)$ and 
  the bifundamentals $\phi$ have charge $(0,2)$.
  The charges of the superpotential should be $(2,2)$.
  Changing all the signs of the building block corresponds to the exchange of $J_{+}$ and $J_{-}$.
  
  We focus on the case where the puncture has one definite sign.
  Based on the form of the anomalies in $\CN=1$ class $\CS$ theories, 
  we propose the following form of the anomalies
    \bea
    \Tr J_{+}
     =     \Tr J_{+}^{3} 
    &=&    q \delta + \sum_{i (\sigma = -1)} \Delta(Y_{i}) + \Gamma, 
           \nonumber \\
    \Tr J_{-}
     =     \Tr J_{-}^{3} 
    &=&    p \delta + \sum_{i (\sigma = +1)} \Delta(Y_{i}) + \Gamma,
           \nonumber \\
    \Tr J_{+}^{2} J_{-}
    &=&    p \omega + \sum_{i (\sigma = +1)} \Omega(Y_{i}) + \Gamma,
           \nonumber \\
    \Tr J_{+} J_{-}^{2}
    &=&    q \omega + \sum_{i (\sigma = -1)} \Omega(Y_{i}) + \Gamma,
           \label{anomalieschiralN=1'}
    \eea
  where the sums are over all the punctures with $\sigma=+$ or $\sigma=-$.
  Also the last terms are
    \bea
    \Gamma
     =   - (3g-3+n)k - \mathfrak{m} (k-1),
           \label{Gamma}
    \eea
  where the first term comes from the gauging
  and the second term comes from the introduction of the discrete $U(1)_{t}$ curvature 
  whose number is $\mathfrak{m} \in \IZ$.
  
  For the theories constructed from the building blocks $\CB^{\sigma, \tilde{\sigma}}_{\sigma^{(b)}}$,
  namely the linear and cyclic quivers, $\CL_{ n-2,\sigma_{tot}}^{(n_{L},\sigma_{L}),(n_{R},\sigma_{R})}$ ($g=0$) 
  and $\CC_{ n, \sigma_{tot} }$ ($g=1$),  
  one can easily check that the relations $\tr J_{\pm} = \tr J_{\pm}^{3}$
  are satisfied because the matters are always of charge $(1,0)$ or $(0,2)$
  (or exchange of $J_{+}$ and $J_{-}$), whose fermionic components are of $(0,-1)$ and $(-1, 1)$.
  
  Some of the parameters in the formulas can be read off by considering the anomalies of the building blocks,
  where $3g-3+n = 0$.
  The anomalies of $\CB^{++}_{+}$ are $\tr J_{-} = - 2 k N^{2}$ and $\tr J_{+} = \tr J_{+}^{2} J_{-}
  = \tr J_{+}J_{-}^{2} = 0$,
  and the anomalies of $\CB^{-+}_{+}$ are $\tr J_{+} = \tr J_{-} = - k N^{2}$ and 
  $\tr J_{+}^{2} J_{-} =  - \tr J_{+} J_{-}^{2} = k N^{2}$,
  by comparing these we get 
    \bea
    \Delta(Y_{max})
     =   - kN^{2}, ~~~
    \delta + \Delta(Y_{min})
     =     0,
    \eea 
  and
    \bea
    \Omega(Y_{max})
     =   - kN^{2}, ~~~
    \omega + \Omega(Y_{min})
     =     2 k N^{2}.
    \eea
  Furthermore, checking the anomalies of the building blocks seen in Section \ref{sec:higgs}
  verifies the second term in \eqref{Gamma}.

  One cannot determine $\delta$ and $\Delta(Y_{min})$ (or $\omega$ and $\Omega(Y_{min})$)
  only from these building blocks.
  However this is enough at least for the theory constructed 
  from the building blocks $\CB^{\sigma \tilde{\sigma}}_{\sigma^{(b)}}$.
  where the number of the pairs-of-pants with $\sigma=+$ ($\sigma=-$) is the same as that of 
  the minimal punctures with $\sigma=+$ ($\sigma=-$):
    \bea
    \Tr J_{+}
     =     \Tr J_{+}^{3} 
    &=&  - k (3g-3+n + n_{-} N^{2}), 
           \nonumber \\
    \Tr J_{-}
     =     \Tr J_{-}^{3} 
    &=&  - k (3g-3+n + n_{+} N^{2}),
           \nonumber \\
    \Tr J_{+}^{2} J_{-}
    &=&    k(2pN^{2} - (3g-3+n) - n_{+} N^{2}),
           \nonumber \\
    \Tr J_{+} J_{-}^{2}
    &=&    k(2qN^{2} - (3g-3+n) - n_{-} N^{2}),
           \label{anomalieschiralN=1quiver}
    \eea
  where $n_{\pm}$ are the numbers of the maximal punctures with sign $\pm$ and $g=0$ or $1$.
  
  Indeed one can check \eqref{anomalieschiralN=1quiver} by explicitly computing the anomalies of 
  $\CL_{ n-2,\sigma_{tot}}^{(n_{L},\sigma_{L}),(n_{R},\sigma_{R})}$ and $\CC_{n, \sigma_{tot}}$.
  Each pair-of-pants with $\sigma=+$ contributes to the anomalies as
  $\tr J_{+} = \tr J_{+}^{2} J_{-} = \tr J_{+} J_{-}^{2} =0$ and $\tr J_{-} = - 2kN^{2}$. 
  The negative sign pair-of-pants has similar anomalies given 
  by the above expressions with an exchange of $J_{+}$ and $J_{-}$.
  Now we consider the vertical bifundamentals, which could appear at the ends of the quiver
  (connecting the flavor $SU(N)$'s)
  or inside the quiver (connecting the gauge $SU(N)$'s).
  Let $n_{\sigma}^{\sigma'}$ be the number of the maximal punctures with sign $\sigma$ 
  attached to the pair-of-pants with sign $\sigma'$.
  Then the contributions from the vertical bifundamentals (at the ends) are given by
  $\tr J_{+} = \tr J_{+} J_{-}^{2} = - \tr J_{-} = - \tr J_{+}^{2} J_{-} = - (n_{-}^{+} - n_{+}^{-})kN^{2}$.
  To calculate the contribution from the vertical bifundamentals inside the quiver, 
  let $N_{+}$ and $N_{-}$ be the numbers of the ``glued punctures'' 
  attached to the pairs-of-pants with $\sigma = \pm$ respectively.
  Then $N_{+} = 3p - n_{min}^{+} - n_{+}^{+} - n_{-}^{+}$
  and $N_{-} = 3p - n_{min}^{-} - n_{+}^{-} - n_{-}^{-}$
  where $n_{min}^{\sigma}$ is the number of the minimal punctures with $\sigma$.
  Note that $p=n_{min}^{+}$ and $q=n_{min}^{-}$.
  Then the net contribution of the vertical inside bifundamentals is computed as
  $\tr J_{+} = \tr J_{+} J_{-}^{2} = - \tr J_{-} = - \tr J_{+}^{2} J_{-} = - \frac{N_{+} - N_{-}}{2}kN^{2}$.
  Finally since the gaugino has $(J_{+},J_{-}) = (1,1)$, the anomalies from the vector multiplets are
  $\tr J_{+} = \tr J_{-} = \tr J_{+}^{2} J_{-} = \tr J_{+} J_{-}^{2} = k (N^{2} - 1)$.
  By summing altogether, we obtain \eqref{anomalieschiralN=1quiver}.

\subsection{Central charges}
\label{subsec:centralcharge}
  The central charges of the $\CN=1$ SCFT are written in terms of the anomalies of the infrared $R_{IR}$ symmetry as
    \bea
    a
     =     \frac{3}{32} \left( 3 \tr R_{IR}^{3} - \tr R_{IR} \right), ~~~
    c
     =     \frac{1}{32} \left( 9 \tr R_{IR}^{3} - 5 \tr R_{IR} \right).
    \eea
  In the case studied in this paper, the IR $R$-symmetry is not necessarily $U(1)_{R}$.
  Instead, $U(1)_{R}$ is  mixed with other $U(1)$ symmetries to give $R_{IR}$ in the infrared.
  To determine this mixing we can use a-maximization \cite{Intriligator:2003jj}, 
  however if there are many $U(1)$ symmetries which can be mixed, the computation could be tedious.
  
  Let us first study two classes of quiver theories
  $\CL_{ n-2,\sigma_{tot}}^{(n_{L},\sigma_{L}),(n_{R},\sigma_{R})}$ and $\CC_{n, \sigma_{tot} }$.
  Indeed these are ``good'' in a sense that the mixing can be understood easily:
  the $U(1)$ symmetries coming from the minimal punctures are baryonic, thus these do not mix with $U(1)_{R}$;
  the $U(1)_{\beta_{i}}$ and $U(1)_{\gamma_{i}}$ are ``periodic'' and there are no preferred directions for them
  so only the overall part of these can be mixed. 
  This is the $U(1)_{t}$ symmetry.
  Therefore in these classes of theories we only need to maximize the trial central charge $a(\epsilon)$
  from the trial $R_{trial}$
    \bea
    R_{trial}
     =     R + \epsilon T
     =     \frac{1}{2} ((1+\epsilon)J_{+} + (1-\epsilon)J_{-}).
    \eea

  Let us compute $a$ and $c$ for a couple of examples.
  First of all, let us consider the building blocks we saw in Section \ref{sec:class}.
  For $\CB_{+}^{++}$ the a-maximization gives $\epsilon= \frac{1}{3}$.
  This is reasonable because at this value of $\epsilon$ all the chiral multiplets have R-charge $\frac{2}{3}$
  in the infrared, which means they are actually free fields.
  Indeed the central charges $a$ and $c$ are
    \bea
    a
     =     \frac{kN^{2}}{24}
     =     2 kN^{2} a_{free}, ~~~
    c
     =     \frac{kN^{2}}{12}
     =     2kN^{2} c_{free} ,
    \eea
  where $a_{free}$ and $c_{free}$ are the central charges of a free chiral multiplet.
  In the same way, we consider other building blocks, $\CB_{+}^{-+}$, $\CB_{+}^{+-}$ and $\CB_{+}^{--}$.
  For all the cases we again obtain $\epsilon = \frac{1}{3}$.

  In the orbifolded SQCD studied in Section \ref{sec:SQCD}, $\CL^{+-}_{2, 0}$, 
  it is easy to compute
    \bea
    \Tr J_{+}
     =     \Tr J_{+}^{3} 
    &=&  - k (N^{2} + 1), ~~~
    \Tr J_{-}
     =     \Tr J_{-}^{3} 
     =   - k (N^{2} + 1),
           \\
    \Tr J_{+}^{2} J_{-}
    &=&    k(N^{2} - 1),~~~
    \Tr J_{+} J_{-}^{2}
     =    k(N^{2} - 1).
    \eea
  Then by maximizing the trial central charge we get $\epsilon = 0$.
  Thus $U(1)_{R}$ is indeed the infrared $U(1)_{R}$ where the bifundametals $Q$'s have charge $1/2$, recovering the known values of SQCD.
  The central charges are given by
    \bea
    a
     =    k \left( \frac{15}{64} N^2- \frac{3}{16} \right), ~~~
    c
     =    k \left( \frac{19}{64} N^2- \frac{1}{8} \right)
    \eea
  giving $k$ times the central charges of $\CN=1$ SQCD with $2N$ flavors.
   
  Next let us consider the theory $\CL^{++}_{2,2}$, which is the orbifold of $\CN=2$ SQCD.
  Again it is straightforward to calculate $\epsilon$ to get $\frac{1}{3}$.
  As in the case of the building blocks, this means that all the chiral multiplets have R-charge $\frac{2}{3}$.
  This is as it should be because the number of the flavors of each gauge group is $N_{f} = 3N$.
  
  We have computed the central charges of the generic linear quiver theories 
  $\CL^{(n_{L},\sigma_{L}), (n_{R},\sigma_{R})}_{n-2, \sigma_{tot}}$ and $\CC_{n, \sigma_{tot}}$.
  For example in the case of the torus, we pick the parametrization $n=p+q$ and $p-q=2k \mathfrak{n}$.
  Then we obtain 
    \bea
    \epsilon
     =     \frac{-p + \mathfrak{n} k+ \sqrt{p^2 - 2 p \mathfrak{n} k + 4 \mathfrak{n}^2 k^2}}{3 k \mathfrak{n}}
    \eea
  which does not depend on the rank of the gauge groups $N$.
  The central charges can be computed easily from this value.
  It is interesting to note that $\epsilon$ always satisfies
    \bea
    - \frac{1}{3} \leq \epsilon \leq \frac{1}{3}.
    \eea
  The bounds are saturated when we set all the signs to be the same, 
  which corresponds to theories obtained from the $\CN=2$ quiver theories by an orbifold.
  In this case all the chiral multiplets become free.
  We have also checked that the ratio of the central charge $a/c$ is in the regime $1/2 < a/c < 3/2$.
  Here we focused on $\CC_{n, \sigma_{tot}}$, 
  however the above results are also satisfied in $\CL^{(n_{L},\sigma_{L}), (n_{R},\sigma_{R})}_{n-2, \sigma_{tot}}$.

\paragraph{Including $U(1)$ curvatures}
  The situation could be different when we study the case with the $U(1)_{\beta}$ or $U(1)_{\gamma}$ curvature.
  Apparently, the symmetric structure of $U(1)_{\beta_{i}}$ or $U(1)_{\gamma_{i}}$ is lost,
  and thus these could mix with $U(1)_{R}$.
  
  The simplest examples are the building blocks in Section \ref{sec:higgs}.
  {\it E.g.}, in the second one from the left in Figure \ref{fig:buildingblock6},
  the $U(1)_{R}$ can mix with all the other $U(1)$'s.
  The computation is tedious thus we quickly see a few results.
  For $k=2$, we can check that the R-charges of all the gauge invariant operators are larger than $\frac{2}{3}$ for arbitrary $N$.
  For $k=3$, one of the R-charges of the gauge invariant operators fails to satisfy the unitarity bound only when $N=2$.
  A similar phenomenon is known to happen for $Y^{p,q}$ theories \cite{Benvenuti:2004:06/064}.
  Presumably there is an accidental symmetry which arises when these fields decouple. 
  It would be interesting to study the condition for having a theory without the unitarity-violating operator.

\section*{Acknowledgements}
We would like to thank 
Antonio Amariti, Luca Mazzucato, Shlomo Razamat, Rak-kyong Seong for useful comments and discussions.
K.~M.~ would like to thank Yuji Tachikawa for useful discussions. 
A.~H.~ would like to thank KIAS, Seoul for their kind hospitality.
The work of K.~M.~ is supported by the EPSRC programme grant ``New Geometric
Structures from String Theory'', EP/K034456/1.




\bibliographystyle{ytphys}
\bibliography{refs}

\end{document}